\documentclass[12pt]{iopart}

\usepackage{graphicx}

\begin{document}

%\review{Tudo o que sempre quis saber sobre a instabilidade de cisalhamento num plasma nao magnetizado} 
\title{Electron-scale shear instabilities: magnetic field generation and particle acceleration in astrophysical jets}

\author{E. P Alves$^1$, T Grismayer$^1$, R A Fonseca$^{1,2}$ and L. O Silva$^1$}
\address{$^1$ GoLP/Instituto de Plasmas e Fus\~ao Nuclear - Laborat\'orio Associado, Instituto Superior T\'ecnico, Lisbon, Portugal}
\address{$^2$ DCTI/ISCTE Instituto Universit\'{a}rio de Lisboa, 1649-026 Lisboa, Portugal}
\eads{\mailto{e.paulo.alves@ist.utl.pt}}
\eads{\mailto{luis.silva@ist.utl.pt}}

\date{\today}

\begin{abstract}

Strong shear flow regions found in astrophysical jets are shown to be important dissipation regions, where the shear flow kinetic energy is converted into electric and magnetic field energy via shear instabilities. The emergence of these self-consistent fields make shear flows significant sites for radiation emission and particle acceleration. We focus on electron-scale instabilities, namely the collisionless, unmagnetized Kelvin-Helmholtz instability (KHI) and a large-scale dc magnetic field generation mechanism on the electron scales. We show that these processes are important candidates to generate magnetic fields in the presence of strong velocity shears, which may naturally originate in energetic matter outburst of active galactic nuclei and gamma-ray bursters. We show that the KHI is robust to density jumps between shearing flows, thus operating in various scenarios with different density contrasts. Multidimensional particle-in-cell (PIC) simulations of the KHI, performed with OSIRIS, reveal the emergence of a strong and large-scale dc magnetic field component, which isÊnot captured by the standard linear fluid theory. This dc component arises from kinetic effects associated with the thermal expansion of electrons of one flow into the other across the shear layer, whilst ions remain unperturbed due to their inertia. The electron expansion forms dc current sheets, which induce a dc magnetic field. Our results indicate that most of the electromagnetic energy developed in the KHI is stored in the dc component, reaching values of equipartition on the order of $10^{-3}$ in the electron time-scale, and persists longer than the proton time-scale. Particle scattering/acceleration in the self generated fields of these shear flow instabilities is also analyzed. 

\end{abstract}

\maketitle

\section{Introduction}

Relativistic jets are found in a wide range of extreme astrophysical scenarios like active galactic nuclei (AGN) and gamma-ray bursts (GRBs) \cite{bridle84,mirabel99}. The energetic outflows of plasma associated with astrophysical jets represent massive sources of free-energy for collisionless plasma instabilities to operate. The onset of plasma instabilities play a central role in dissipating the jet's kinetic energy into electric and magnetic turbulence \cite{gruzinovwaxman99,medvedev99} resulting in particle acceleration to ultra-high energies and nonthermal radiation emission. A deep understanding of these processes and their interplay is challenging, requiring full kinetic simulations to address their highly nonlinear nature. First principle modeling of these processes are, however, computationally intensive due to the wide range of temporal and spatial scales involved. Therefore, full kinetic simulations demand massive computational resources and advanced numerical and visualization techniques.

Much attention has been devoted to relativistic shocks, which are thought to be a strong mechanism for particle acceleration. Such shocks arise from the collision and bulk interpenetration of different velocity plasma shells, due to either intermittencies or inhomogeneities of the ejecta. The Weibel \cite{weibel59} and the purely transverse two stream instabilities \cite{silva03} act as the dissipation mechanism in these scenarios, and are critical for shock formation. A vast number of fully kinetic simulations have focused on shock formation settings, where long-lived equipartition magnetic field generation via the Weibel instability has been observed \cite{silva03,fonseca03,frederiksen04,nishikawa05}. A Fermi-like particle acceleration process has also been identified in simulations of long-term evolution of collisionless shocks \cite{spitkovsky08,martins09}. These previous works have only considered shearless flows. 

However, in addition to bulk plasma collision sites, the transition layers of shear flows have also been probed \cite{gruzinov08} and shown to constitute important dissipation regions \cite{alves12,grismayer13,Liang13}. Increasing evidence has pointed to a general stratified organization of the structure of jets in AGN and GRBs \cite{granot03, rieger04}, where different internal shear layers can occur; rotating inner cores vs. axially moving outer shells, or fast inner cores vs. slower outer shells. Moreover, external shear layers, resulting from the interaction of the jet with the interstellar medium, may also be considered. In these scenarios, collisionless shear instabilities such as the Kelvin-Helmholtz instability (KHI) \cite{dangelo65,gruzinov08,macfadyen09} play a role in the dissipation of the jet kinetic energy into electric and magnetic turbulence \cite{macfadyen09,alves12,Liang13}. In fact, the combined effect of shear flow with collisionless shock formation has not yet been addressed, and may also lead to interesting novel phenomenology since density inhomogeneities generated by shear instabilities can also constitute important scattering sites for particle acceleration. Recent fully kinetic simulations of shear flow settings have probed the self-consistent evolution of the electron-scale KHI, demonstrating that the operation of kinetic effects are responsible for the generation of large-scale, equipartition magnetic fields \cite{alves12,grismayer13}. Nonthermal particle acceleration has also been investigated in hybrid electron-positron-ion shear flows \cite{boettcher,Liang13}, with different pair/ion ratio compositions, showing spectral features similar to those found in GRBs.

In laboratory experiments, scenarios where the unmagnetized KHI can be triggered are now being examined both in the collisional \cite{harding09,Hurricane12} and in the collisionless regimes\cite{kuramitsu12} (the latter is explored in this paper).
In this work we focus on electron-scale processes triggered by velocity shears, namely the unmagnetized KHI and a dc magnetic field generation mechanism. In Section 2, we develop the linear theory for the cold unmagnetized KHI, and analyze the impact of density contrast between sharp shearing flows. We find the onset of the KHI is robust to density contrasts, allowing for a strong development in various density contrast regimes (inner shears with low density contrasts, and outer shears with high density contrasts). We then extend the analysis to finite shear gradients, where we find that KHI growth rate decreases with increasing shear gradient length. Particle-in-cell (PIC) simulations are performed to verify the theoretical predictions. At late times, PIC simulations reveal the formation of a large-scale, dc magnetic field extending along the entire shear surface between flows, which is not predicted by the linear KHI theory. This dc magnetic field is the dominant feature of the magnetic field structure of the instability at late times. In Section 3, we find that the dc magnetic field results from kinetic effects associated with electron mixing between shearing flows, which is driven by the nonlinear development of the cold KHI. The dc magnetic field generation is discussed and an analytical model is developed that captures the main features of the dc magnetic field evolution and saturation. In Section 4, we analyze the dynamics of the electrons in the self-generated fields. The electrons are scattered in the self consistent electric and magnetic fields generated by the KHI, and are accelerated to high energies. We discuss the particle energy spectra resulting from the development of these shear instabilities, and we investigate the mechanism underlying the acceleration of energetic particles using advanced particle tracking diagnostics.

\newpage
\section{The cold, unmagnetized, electron-scale KHI}
\label{sec:fluid_regime}

The KHI is a well known instability that is driven by velocity shear. This instability was first derived for neutral shearing fluids within the hydrodynamic framework, where the flows interact via pressure gradients \cite{Chan61,DrazinReid81}. The KHI in charged fluids (plasmas) has also been studied within the MHD framework \cite{dangelo65,ThomasWinske91}, where the shearing flows also interact via electric and magnetic fields, in addition to pressure gradients. In both these frameworks, the length (and time) scales involved are much larger than the kinetic scales associated with the particles that make up the neutral or charged fluid. In this work, we study the KHI undergone by the electron fluid component of the plasma; the ion dynamics, due to their large inertia, are neglected and are assumed to be unperturbed during the development of the electron-scale KHI. The physics underlying the development of the KHI at the electron-scale is different from the more usual hydrodynamics and MHD forms of the instability, and leads to interesting features that are not observed in more macroscopic frameworks.
It is important to note that the KHI can occur at various scales (from electron-kinetic to MHD scales), and that the cross-scale connection and interplay of these instabilities remains to be understood.
In this Section, we present the linear two-fluid theory of the electron-scale KHI for an initially cold and unmagnetized plasma shear flow. We generalize for arbitrary velocity and density profiles, and derive analytical solutions for step-like velocity shear and density profiles. The theoretical results are then compared and verified with PIC simulations.

\subsection{Linear two-fluid theory}
\label{sec:lin_develop_khi}

\begin{figure}[t]
\centering
	\includegraphics{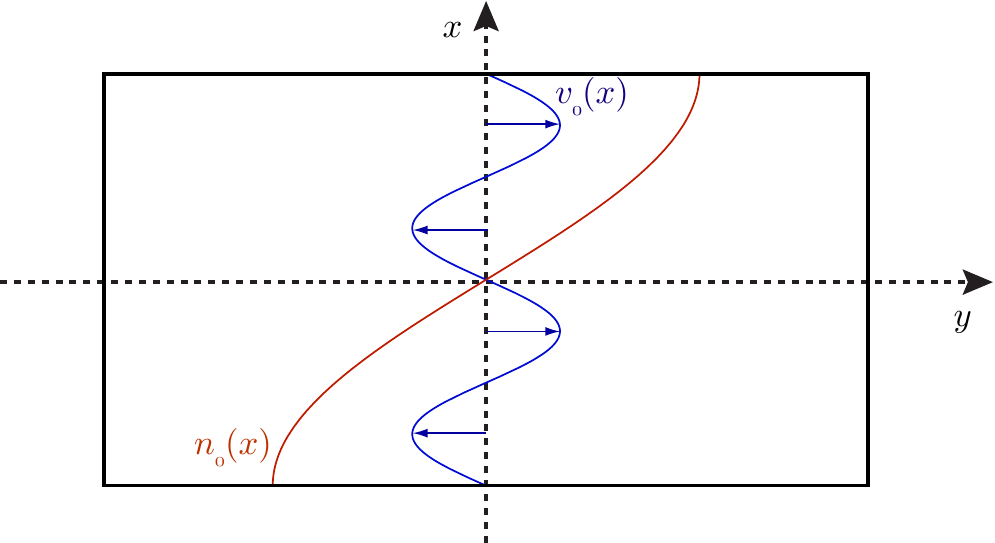}
	\caption{Theoretical setting for a 2D shear flow, with arbitrary velocity and density profiles $v_0(x)$ and $n_0(x)$, respectively}
	\label{fig:generalcase}
\end{figure}

In the case of an initially unmagnetized plasma in equilibrium, there is no need for a pressure term to balance out the magnetic force. The equilibrium of the system is then naturally obtained by taking the cold limit of the plasma. In order to describe the linear regime of the electron KHI, we employ the relativistic fluid theory of plasmas. The equations that constitute this theoretical framework are
\begin{equation} \label{eq:cont}
		\frac{\partial \rho}{\partial t} + \mathbf{\nabla} \cdot \mathbf{J} = 0,
	\end{equation}

	\begin{equation} \label{eq:motion}
		\frac{\partial \mathbf{p}}{\partial t} + \left( \mathbf{v} \cdot \mathbf{\nabla} \right) \mathbf{p} = - e \left( \mathbf{E} + \frac{\mathbf{p}}{\gamma m_e} \times \mathbf{B} \right),
	\end{equation}
	\begin{equation} \label{eq:max1}
		\mathbf{\nabla} \times \mathbf{E} = - \frac{\partial \mathbf{B}}{\partial t},
	\end{equation}
	\begin{equation} \label{eq:max2}
 		c^2 \mathbf{\nabla} \times \mathbf{B} = - \frac{1}{\epsilon_0} \mathbf{J} + \frac{\partial \mathbf{E}}{\partial t}.
	\end{equation}
The equations are written in SI units. Eq.~(\ref{eq:cont}) and Eq.~(\ref{eq:motion}) are respectively the continuity and conservation of momentum equations. Eq.~(\ref{eq:max1}) is Faraday's equation and Eq.~(\ref{eq:max2}) is Ampere's equation. Here, $\rho=en$ where $n$ is the plasma density, $\mathbf{J}$, $\mathbf{E}$ and $\mathbf{B}$ are the current density, electric field and magnetic field vectors, respectively. $\mathbf{p} = \gamma m_e \mathbf{v}$ and $\mathbf{v}$ are the linear momentum and velocity vectors, where $\gamma = \left( 1- v^2/c^2 \right)^{-1/2}$ is the relativistic Lorentz factor; $c$ is the speed of light, $m_e$ and $e$ are, respectively, the electron mass and electron charge, and $\epsilon_0$ is the electric permittivity of vacuum. We assume a two-dimensional (2D) cold relativistic shear flow with initial velocity and density profiles described by,
\begin{equation}
\mathbf{v}=\left ( 0, v_0 \left ( x \right ), 0 \right ), \qquad n=n_0 \left ( x \right ),
\end{equation}
respectively (Figure \ref{fig:generalcase}).
Due to the 2D assumption, the system lies in the $xy$ plane and sustains electric and magnetic fields of the form:
	\begin{equation}
	\label{eq:fieldstructure}
		\mathbf{E}=\left ( E_{x} \left ( x, y ,t\right ), E_{y} \left ( x, y, t \right ), 0 \right ) \qquad \mathbf{B}=\left ( 0, 0, B_{z} \left ( x, y,t  \right ) \right )
	\end{equation}
Since we are first interested in the linear evolution of the system, we linearize all physical quantities:
	\begin{equation} \label{eq:lin_quants}
	\cases{
			n(x,y,t)=	n_0(x) + n_1(x,y,t) \\
			\mathbf{v}(x,y,t)=	v_0(x) \mathbf{e_y} + \mathbf{v_1}(x,y,t) \\
			\mathbf{E}(x,y,t)=	\mathbf{E_1}(x,y,t) \\
			\mathbf{B}(x,y,t)=	\mathbf{B_1}(x,y,t) \\			
			\mathbf{J}(x,y,t)=	\mathbf{J_1}(x,y,t) \\						
		}
	\end{equation}
The subscripts $0$ and $1$ denote zeroth and first-order quantities, respectively. External electric and magnetic fields are absent and therefore the zeroth-order quantities of these fields are zero. Since the structures produced by the instability emerge along the $y$ direction, we look for solutions of the form:
	\begin{equation}
		\label{eq:sol_type}
		Q_1(x,y,t)= Q_1(x)e^{i\left( k y - \omega t \right)}
	\end{equation}
The ions are assumed to be infinitely massive and thus free streaming, and consider only perturbations in the electron dynamics. The linearized equation of continuity for the electron fluid reads
	\begin{equation}
		\label{eq:cont1}
		\frac{\partial}{\partial t} n_{1} + \mathbf{\nabla} \left (  n_0 \mathbf{v_{1}}\right ) + \mathbf{\nabla} \left (  n_{1} \mathbf{v_0}\right ) = 0,
	\end{equation}
Substituting the solution form of Eq.~(\ref{eq:sol_type}) into $n_1$ and $v_1$, we arrive at:
	\begin{equation}
		\label{eq:cont2}
		n_{1}=\frac{-i}{\omega - k v_0} \left(
								\frac{\partial}{\partial x} \left(  v_{x1} n_0\right) + i k n_0 v_{y1}
								\right).
	\end{equation}
The linearized equation of motion of the electrons is given by
	\begin{equation}
		\label{eq:motion1}
		\frac{\partial}{\partial t} \mathbf{p_1} + \left( \mathbf{v_1} \cdot \mathbf{\nabla} \right) \mathbf{p_0} + \left( \mathbf{v_0} \cdot \mathbf{\nabla} \right) \mathbf{p_1} =
		 - e \left ( \mathbf{E_1} + \mathbf{v_0} \times \mathbf{B_1} \right ),
	\end{equation}
The zeroth and first-order momentum are, respectively,
	\begin{equation}
		\label{eq:relmomvel}
		 \mathbf{p_0} = \gamma_0 m_e \mathbf{v_0},
		 \qquad 
		 \mathbf{p_1} = \left . \left(\mathbf{v_1} \cdot \mathbf{\nabla} \right) \mathbf{p} \: \right | _{v=v_0} = m_e \mathbf{v_1} \gamma_0 + m_e \gamma_0^3 \frac{\mathbf{v_1} \cdot \mathbf{v_0}}{c^2}\mathbf{v_0}
	\end{equation}
Inserting Eq.~(\ref{eq:relmomvel}) into Eq.~(\ref{eq:motion1}) and solving for $\mathbf{v_1}$, we arrive at
	\begin{eqnarray}
		\label{eq:v1x}
		v_{x1} &=& \frac{1}{\gamma_0 m_e} \frac{-ie}{\omega - k v_0} \left( E_{x1} + v_0 B_{z1} \right) \\
		\label{eq:v1y}
		v_{y1} &=& \frac{1}{\gamma_0^3 m_e} \frac{-i}{\omega - k v_0} \left( e E_{y1} + m_e v_{x1} \frac{\partial}{\partial x} \left( \gamma_0 v_0\right)  \right).
	\end{eqnarray}
Combining Eq.~(\ref{eq:v1x}) and Eq.~(\ref{eq:v1y}) with Eq.~(\ref{eq:cont2}) we compute the perturbed current density $\mathbf{J_1}= -e \left( n_0 \mathbf{v_1} +n_1 \mathbf{v_0} \right)$,
	\begin{eqnarray}
	\label{eq:curr_x}
		J_{x1}  &=& \frac {i e^2 n_0}{\gamma_0 m_e} \frac {1}{\omega-k v_0} \left ( E_{x1} + v_0 B_{z1} \right ) \\
	\label{eq:curr_y}
		J_{y1}  &=& \frac {i e^2 n_0}{\gamma_0m_e} \frac {\omega}{\left ( \omega-k v_0 \right ) ^2} E_{x1} + 
		\frac{\partial}{\partial x} \left ( \frac {e^2 n_0}{\gamma_0m_e} \frac {v_0}{\left ( \omega-k v_0 \right ) ^2} \left ( E_{x1} + v_0 B_{z1} \right ) \right ).
	\end{eqnarray}
We now couple these current densities to Maxwell's equations in order to close our system of equations. The linearized form of Eq.~(\ref{eq:max1}) and Eq.~(\ref{eq:max2}) are written as:
	\begin{equation}
		\label{eq:maxwell1}
		\mathbf{\nabla} \times \mathbf{E_1} = - \frac{\partial \mathbf{B_1}}{\partial t},
	\end{equation}
	\begin{equation}
		\label{eq:maxwell2}
 		c^2 \mathbf{\nabla} \times \mathbf{B_1} = - \frac{1}{\epsilon_0} \mathbf{J_1} + \frac{\partial \mathbf{E_1}}{\partial t},
	\end{equation}
These two equations, Eq.~\ref{eq:maxwell1} and Eq.~\ref{eq:maxwell2} are combined by taking the curl of Eq.~(\ref{eq:maxwell1}) and substituting in Eq.~(\ref{eq:maxwell2}),
	\begin{equation}
		\label{eq:maxwell3}
		\mathbf{\nabla} \times \left( \mathbf{\nabla} \times \mathbf{E_1} \right) = - \frac{1}{c^2} \left( \frac{1}{\epsilon_0} \frac{\partial \mathbf{J_1}}{\partial t} + \frac{\partial^2 \mathbf{E_1}}{\partial t^2} \right),
	\end{equation}
Splitting Eq.~(\ref{eq:maxwell3}) into its components and inserting the candidate plane-wave solutions of the form of Eq.~(\ref{eq:sol_type}), we obtain
	\begin{eqnarray}
		\label{eq:maxwell4}
		&i k& \frac{\partial E_{y1}}{\partial x} = i \frac{\omega}{c^2 \epsilon_0} J_{x1} + \left( \frac{\omega^2}{c^2} -k^2 \right) E_{x1} \\
		\label{eq:maxwell5}
		&i k &\frac{\partial E_{x1}}{\partial x} - \frac{\partial^2 E_{y1}}{\partial x^2}= i \frac{\omega}{c^2 \epsilon_0} J_{y1} + \frac{\omega^2}{c^2} E_{y1}.
	\end{eqnarray}
Next, we insert the current densities, Eq.~(\ref{eq:curr_x}) and Eq.~(\ref{eq:curr_y}) into the equations Eq.~(\ref{eq:maxwell4}) and Eq.~(\ref{eq:maxwell5}) which, after some algebra, leads to the following equation describing the linear electromagnetic eigenmodes of the system:
	\begin{equation}
		\label{eq:eigenmodes}
		\frac{\partial}{\partial x} \left [A \frac{\partial E_{y1}}{\partial x} \right ] +B \frac{\partial E_{y1}}{\partial x} + C E_{y1} = 0
	\end{equation}
where the functions $A$, $B$, and $C$ are:
	\begin{equation}
		\cases{
			A=	\frac {\omega ^2}{c^2}
				\left(  \frac{1}{\gamma_0^2} \frac {\omega_p ^2}{\left( \omega -k v_0 \right) ^2} -1 \right)
				\left ( \frac {\omega ^2}{c^2} - \frac {\omega_p ^2}{c^2} - k^2 \right )	\\		
			B= 2 \frac {\omega ^2}{c^2}
				\left( \frac{1}{\gamma_0^2} \frac {\omega_p ^2}{\left( \omega -k v_0 \right) ^2} -1 \right)
				\left ( \frac{\partial}{\partial x} \frac {\omega_p ^2}{c^2} \right ) \qquad \\		
			C=	\frac {\omega ^2}{c^2}
				\left(  \frac{1}{\gamma_0^2} \frac {\omega_p ^2}{\left( \omega -k v_0 \right) ^2} -1 \right)
				\left ( \frac {\omega ^2}{c^2} - \frac {\omega_p ^2}{c^2} - k^2 \right )^2
		}
	\end{equation}
Here, $\omega_p = \sqrt{n_0 e^2/\gamma_0\epsilon_0 m_e}$ denotes the relativistic electron plasma frequency (which is a function of $x$ as it depends on the plasma density profile $n_0(x)$). For general density and velocity fields, Eq.~(\ref{eq:eigenmodes}) may only be solved numerically. However, analytical solutions may be obtained for special settings where Eq.~(\ref{eq:eigenmodes}) is simplified. We now derive an analytical solution of Eq.~(\ref{eq:eigenmodes}) for such a setting.

\subsubsection{Step velocity shear and density profiles} 
\label{sec:disp_rel_analysis}
\begin{figure}[t!]
\centering
	\includegraphics{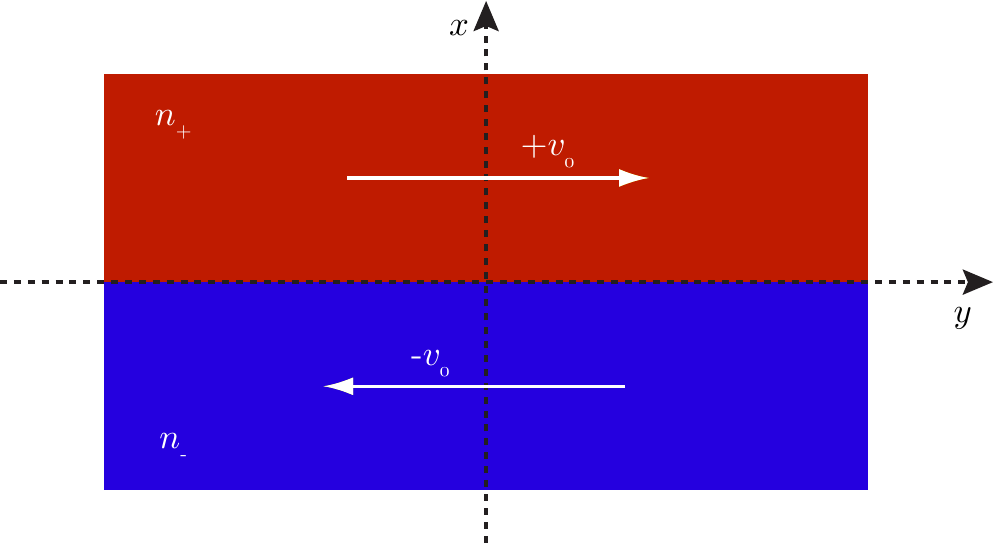}
	\caption{Simplified theoretical setting: tangential discontinuity velocity shear between different uniform density flows.}
	\label{fig:special_case}
	\end{figure}

We consider the following step-function velocity shear profile,
	\begin{equation}
		\label{eq:v_jump}
		\vec{v_0}(x) = 
			\cases{ 
				   +v_0~\vec{e_y}  & $x > 0$ \\
				   -v_0~\vec{e_y}  & $x < 0$
			}
	\end{equation}
and step-function density profile,
	\begin{equation}
		\label{eq:n_jump}
		n_0(x) = 
			\cases{ 
				   n_{+}  & $x > 0$ \\
				   n_{-}  & $x < 0$
			}.
	\end{equation}
The values $v_0$ and $n_\pm$ are constants. This setting translates into two counter propagating flows with different densities which shear at the plane $x=0$ (Figure \ref{fig:special_case}) and generalizes the standard configuration of equal density flows. 
Inserting these profiles into Eq.~(\ref{eq:eigenmodes}), we note that the functions $A$ and $C$ are step-like functions, and that the function $B$ is proportional to $\delta (x)$ since it contains the derivative of the density profile (embedded in the plasma frequency, $\omega_p$). We begin by integrating Eq.~(\ref{eq:eigenmodes}) for $x > 0 $ and $x < 0$ separately and later join the two solutions at the discontinuity plane $x = 0$.
Applying the well known dielectric boundary conditions to our system, we deduce that $E_y$, being the component of the electric field tangential to the dielectric interface, must be continuous, i.e., $E_{y1}(0^+)=E_{y1}(0^-)\equiv E_{y1}(0)$. Thus, for $x \neq 0$, the functions $A$ and $C$ are constants and $B=0$, leading evanescent wave solutions:
	\begin{equation}
		\label{eq:solutions1}
		E_{y1}(x) = E_{y1}(0) e^{-k_{\perp}|x|}  
	\end{equation}
where $k_{\perp}=\sqrt{k^2 + \omega_{p+}^2/c^2 - \omega^2/c^2}$ and $\omega_{p\pm}$ is the electron plasma frequency of the $n_\pm$ plasma. The dispersion relation is finally deduced from the derivative-jump of the electric field at the discontinuity plane. To obtain the derivative jump-condition we perform the standard procedure of integrating Eq.~(\ref{eq:eigenmodes}) over the interval $-\epsilon < x < \epsilon$, and then take the limit $\epsilon \rightarrow 0$. The first term of Eq.~(\ref{eq:eigenmodes}) is trivially integrated and the integration of the third term yields $0$ in the limit $\epsilon \rightarrow 0$. The second term, however, is the product between a Heaviside step-function and a Dirac delta function $\delta(x)$, and is to be evaluated as follows:
	\begin{equation}
		\label{eq:prod_delta_step}
		\lim_{\epsilon \to 0} \int_{-\epsilon}^{+\epsilon} f_\mathrm{step}(x)  \delta(x) \, dx = \frac{f_\mathrm{step}(0^+) + f_\mathrm{step}(0^-)}{2}
	\end{equation}
The derivative jump-condition is thus given by,
	\begin{eqnarray}
		\label{eq:deriv_jump_cond}
		\frac{\partial E_{y1}}{\partial x}(0^+) \left( 
										\frac{\omega_{p+}^2-\omega_{p-}^2}{c^2} \frac {\omega ^2}{c^2}
										\left( \frac{1}{\gamma_0^2} \frac {\omega_{p+}^2}{\left( \omega -k v_0 \right) ^2} -1 \right)
										+A(0^+)
									\right) + \nonumber \\
		\frac{\partial E_{y1}}{\partial x}(0^-) \left( 
										\frac{\omega_{p+}^2-\omega_{p-}^2}{c^2} \frac {\omega ^2}{c^2}
										\left( \frac{1}{\gamma_0^2} \frac {\omega_{p-}^2}{\left( \omega -k v_0 \right) ^2} -1 \right)
										- A(0^-)
									\right) = 0						
	\end{eqnarray}
Finally, manipulating Eq.~(\ref{eq:deriv_jump_cond}) we obtain the following dispersion relation
	\begin{eqnarray}
		\label{eq:disp_rel}
		\sqrt{\frac{n_-}{n_+}+\frac{k'^2}{\beta_0^2}-\omega'^2} 
			\left[
				\left (\omega'+k' \right)^2 - \left(\omega'^2-k'^2\right)^2
			\right] + \nonumber \\ 
		\sqrt{1+\frac{k'^2}{\beta_0^2}-\omega'^2} 
			\left[
				\frac{n_-}{n_+} \left (\omega'-k' \right)^2 - \left(\omega'^2-k'^2\right)^2
			\right] = 0,
\end{eqnarray}
where $\beta_0=v_0/c$, $\omega' = \gamma_0\omega/\omega_{p+}$ and $k' = \gamma_0k v_0/\omega_{p+}$ (where $\omega_{p\pm}=\sqrt{n_\pm e^2/\gamma_0\epsilon_0 m_e}$) are respectively the normalized frequency and wave number in the dispersion relation. Although this model contains ideal velocity and density profiles, it is a useful tool to provide insights into the behaviour of the KHI in the presence of density contrasts between shearing flows.

\begin{figure}[t!]
\centering
	\includegraphics{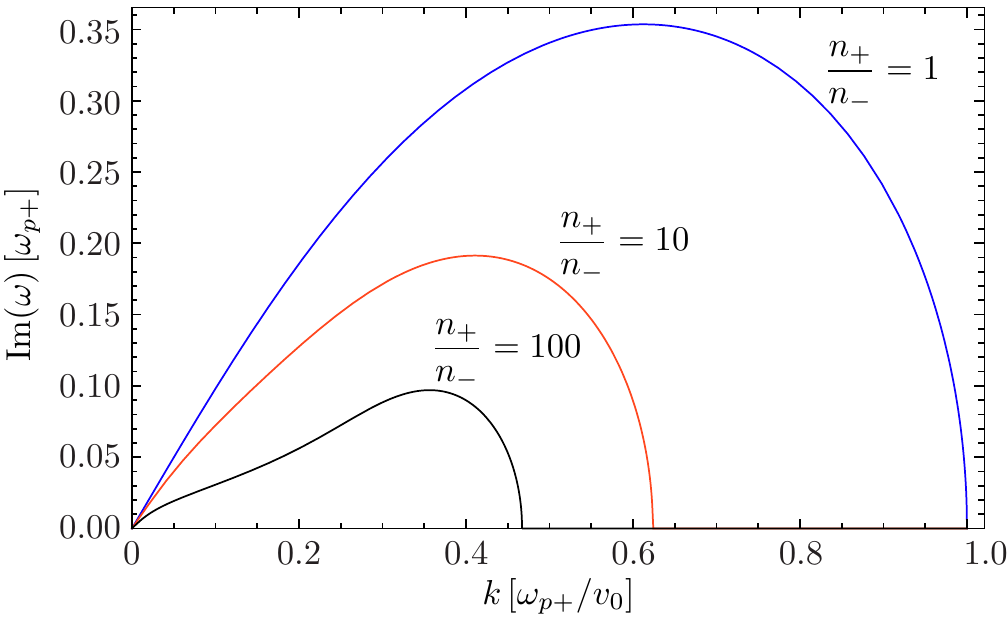}
	\caption{Growth rate of unstable modes for the values $n_+/n_-=1, 10 , 100$.}
	\label{fig:unstable_modes}
\end{figure}

The density contrast is embedded in the dispersion relation through the density ratio, $n_+/n_-$. In the density symmetric limit, $n_+/n_-=1$, Eq.~(\ref{eq:disp_rel}) reduces to a biquadratic equation in $\omega'$, and we recover the analytical solution presented in \cite{gruzinov08}:
	\begin{equation}
		\label{eq:disp_rel_gruzinov}
		\Gamma'=\Im(\omega') = \sqrt{\frac{1}{2} \left( \sqrt{1 + 8 k'^2} -1- 2 k'^2\right) }
	\end{equation}
Eq.~(\ref{eq:disp_rel_gruzinov}) gives the growth rate of the unstable  modes, and is plotted in Figure \ref{fig:unstable_modes}. If we develop at the first order in $k'$ the dispersion relation Eq.~(\ref{eq:disp_rel_gruzinov}), one obtains
\begin{equation}
\Gamma'\simeq k'
\end{equation}
which corresponds to the KHI dispersion relation obtained in the ideal hydrodynamic model for a symmetric shear flow in the absence of surface tension and gravity \cite{Chan61}. The two fluids plasma model dispersion relation differs here from the classical hydrodynamics results by introducing a cut-off at $k'=1$. There is, therefore, a maximum value of the curve that corresponds to the growth rate ($\Gamma'_\mathrm{max}$) of the fastest growing mode ($k'_\mathrm{max}$). These quantities satisfy $\partial_{k'}\Gamma=0$ and are given by:
	\begin{equation}
		\label{eq:gr_gruzinov}
		\Gamma_\mathrm{max}' = \Im(\omega_\mathrm{max}') = \frac{1}{\sqrt{8}}
	\end{equation}
and
	\begin{equation}
		\label{eq:kmax_gruzinov}
		k_\mathrm{max}' =\sqrt{ \frac{3}{8} }
	\end{equation}

The real part of $\omega'$ vanishes over the range of unstable modes meaning that the unstable modes are purely growing waves which is consistent with the symmetry of the system. Note that these electron-scale unstables modes occur when the plasma is considered to be cold, i.e., $v_\mathrm{th} \ll v_0$, whereas compressible MHD or Hydro modes in an initially unmagnetized plasma are only unstable for $v_0 < \sqrt{2}c_s= \sqrt{2} v_\mathrm{th}\sqrt{m_e/m_i}$, which correspond to very slow (or very hot) flows \cite{miura82}. Therefore, shear flow instabilities in initially unmangetized conditions with fast drift velocities (relative to the temperature) can only develop on the electron-scale.

\begin{figure}[t]
\centering
	\includegraphics{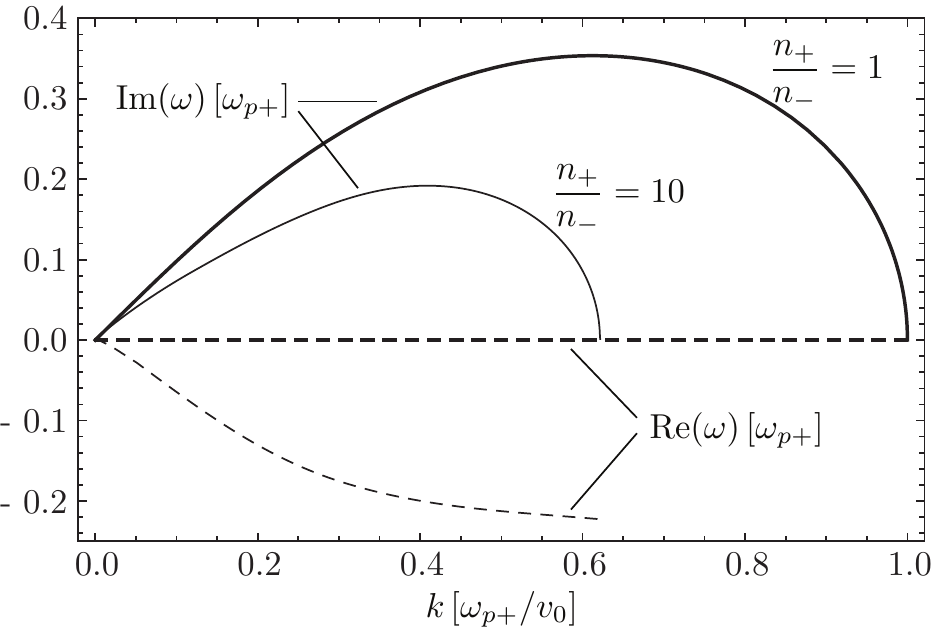}
	\caption{Real (dashed curves) and imaginary (solid curves) parts of $\omega'$ for the symmetric ($n_+/n_-=1$) and an asymmetric ($n_+/n_-=10$) density regimes. The density symmetric and asymmetric regimes are represented by the blue and red curves, respectively.}
	\label{fig:stand_prop}
\end{figure}

In the case of a density jump ($n_+/n_- > 1$), Eq.~(\ref{eq:disp_rel}) has to be solved numerically. Figure \ref{fig:unstable_modes} illustrates the effect of the density asymmetry on the growth rate of the unstable modes for multiple values of $n_+/n_-$. The values of the density ratio are changed assuming $n_+$ fixed so that the normalizing frequency, $\omega_{p+}/ \gamma_0$, and wave number, $\omega_{p+}/(v_0 \gamma_0)$, which determine the axes scales of Figure \ref{fig:unstable_modes}, remain constant. We also consider that $n_+$ corresponds to the denser flow. Thus, larger density ratios are achieved by lowering the value of $n_-$. The qualitative evolution of the unstable modes is independent of the value of $n_+/n_-$, indicating that the general features of the instability are maintained. When $n_+/n_- > 1$ the frequency $\omega'$ acquires a real part over the range of unstable modes leading to propagation (Figure \ref{fig:stand_prop}). The drifting character of the unstable modes results from an unbalanced interaction when each flow has different densities. The dispersion relation Eq.~(\ref{eq:disp_rel}) in the small $k$ limit reduces to
\begin{equation}
\omega' = \frac{k'}{1+\sqrt{r}}\left((\sqrt{r}-1)+2ir^{1/4}\right),
\end{equation}
where $r=\sqrt{n_-/n_+}$. This asymptotic result can be verified in Figure \ref{fig:unstable_modes} and Figure \ref{fig:stand_prop}. However this result do not coincide with the dispersion relation obtained in the ideal hydrodynamics model when there is a density jump, $\Gamma_\mathrm{hydro}=2k\sqrt{r}/(1+r)$. As we noticed before the two growth rates only coincide in the small $k$ limit when $r=1$. 
 In the regime $n_+/n_->1$, the unstable oscillations develop differently in each flow due to their different densities. The growing oscillations are more strongly manifested in the lower density flow ($n_-$) and will thus drift in the direction of the $n_-$ bulk flow. On the other hand, in the density symmetric regime, the surface interaction between flows is balanced; the unstable modes develop equally in each flow, leading to the development of purely growing waves, as previously discussed. The typical growth rate of the KHI ($\Gamma_\mathrm{max}/\omega_{p+}$), as was observed in Figure \ref{fig:unstable_modes}, slows down as $n_+/n_-$ increases. This is because the shear surface current sheets decrease as $n_-$ is lowered. In the limit $n_- \rightarrow 0$ ($n_+/n_- \rightarrow \infty$), we obtain a free streaming plasma in vacuum where the development of the KHI is inhibited, $\Gamma_\mathrm{max}/\omega_{p+} \rightarrow 0$, as expected. 
\begin{figure}[t]
\centering
	\includegraphics{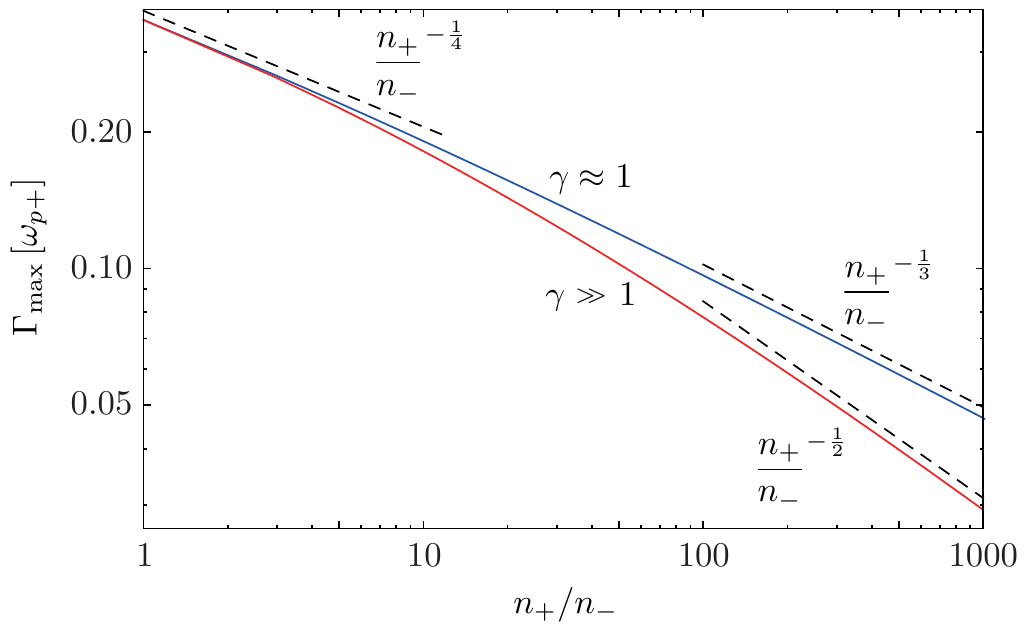}
	\caption[Scalings of the KHI with density contrast between flows.]{Scalings of the growth rate of the KHI ($\Gamma_\mathrm{max}/\omega_{p+}$) with the density ratio between shearing flows. The blue and red curves characterize non-relativistic ($\gamma \approx 1$) and highly-relativistic ($\gamma \gg 1$) settings, respectively.}
	\label{fig:gr_scalings}
\end{figure}
The scaling relations of the KHI with $n_+/n_-$ are shown in Figure \ref{fig:gr_scalings}. In the similar density regime, $n_+/n_- \approx 1$, the growth rate scales as $\Gamma_\mathrm{max}/\omega_{p+} \propto \left(n_+/n_-\right)^{-1/4}$ for both relativistic and non-relativistic shears. In the high density contrast regime, $n_+/n_- \gg 1$, the growth rate scales as $\Gamma_\mathrm{max}/\omega_{p+} \propto \left(n_+/n_-\right)^{-1/3}$ for non-relativistic shears, and  $\Gamma_\mathrm{max}/\omega_{p+} \propto \left(n_+/n_-\right)^{-1/2}$ for highly-relativistic shears. Note also that in the $n_+ \rightarrow 0$ limit ($\omega_p^+ \rightarrow 0$), a scenario where the $n_-$ plasma streams in vacuum, the KHI shuts down. At large $n_+/n_-$ regimes, the KHI dominates over other common plasma instabilities in unmagnetized scenario such as the Weibel and Two-Stream instabilities. The growth rates of the Weibel \cite{silva02} and Two-Stream instabilities \cite{oneil79} scale as $\Gamma_\mathrm{Weibel}/\omega_{p+} \propto \left(n_+/n_-\right)^{-1/2}$ and $\Gamma_\mathrm{2-stream}/\omega_{p+} \propto \left(n_+/n_-\right)^{-1/3}$, respectively. Both growth rates decay more rapidly with $n_+/n_-$ than the growth rate of the KHI. The physics of large density-contrast settings will thus be mainly determined by the evolution of the KHI. Hence, in realistic astrophysical settings with high density contrasts, where various plasma instabilities are triggered simultaneously, magnetic field generation can also be attributed to the development of the KHI.

\subsubsection{Effect of mobile ions}

The influence of mobile ions on the theory previously shown is easily incorporated. The ion fluid obeys the same equations as the electron fluid and only the charge and the mass of the ions are the physical parameters that could impact the dispersion relation. For the sake of clarity, we will restrict ourselves to initial equal density plasma. Following the exact same derivation as aforesaid, we obtain a differential equation with the same form of Eq.(\ref{eq:eigenmodes}) where the plasma frequency needs to be renormalized, $\omega_p^2\rightarrow \omega_p^2(1+m_e/m_i)$. In the case of heavy ions, $m_i\gg m_e$ the effect can be considered negligible. On the other hand, for an electron-positron plasma, the transverse wavenumber $k_{\perp}$ (see Eq.(\ref{eq:solutions1})) is rescaled with the plasma frequency (that is multipled by $\sqrt{2}$) and so is the wave number $k_{\mathrm{max}}'$ associated to the maximum growth rate that peaks at $\Gamma_{\mathrm{max}}'=1/2$.

\subsection{Comparisons with PIC simulations}

Numerical simulations were performed with OSIRIS \cite{fonseca03,fonseca08}, a fully relativistic, electromagnetic, and massively parallel PIC code. We have simulated 2D systems of shearing slabs of cold ($v_{0} \gg v_{th}$, where $v_{th}=10^{-3}c$ is the thermal velocity) unmagnetized electron-proton plasmas with a realistic mass ratio $m_p/m_e=1836$ ($m_p$ is the proton mass), and evolve it until the electromagnetic energy saturates on the electron time scale. We explored a subrelativistic shear flow scenario with $v_0 = 0.2 c$. The setup of the numerical simulations is prepared as follows. The shear flow initial condition is set by a velocity field with $v_0$ pointing in the positive $x_1$ direction, in the upper and lower quarters of the simulation box, and a symmetric velocity field with $-v_0$ pointing in the negative $x_1$ direction, in the middle-half of the box. Note that the coordinates $(x_1,x_2,x_3)$ used in the PIC simulations correspond to the cartesian coordinates $(y,x,-z)$ of the theory presented in the previous sections. Initially, the systems are charge and current neutral, and the shearing flows have equal densities. The simulation box dimensions are $10 \times 10  ~ ( c/\omega_p )^2$, where $\omega_p = ( ne^2/\epsilon_0 m_e)^{1/2}$ is the plasma frequency, and we use $20$ cells per electron skin depth ($c/\omega_p$) in the longitudinal direction and  $200$ cells per electron skin depth ($c/\omega_p$) in the transverse direction. Periodic boundary conditions are imposed in every direction and we use 36 particles per cell.  In order to ensure result convergence, higher numerical resolutions and more particles per cell were tested.

\subsubsection{Equal density shear flows}

\begin{figure}[t!]
\centering
\includegraphics[width=\columnwidth]{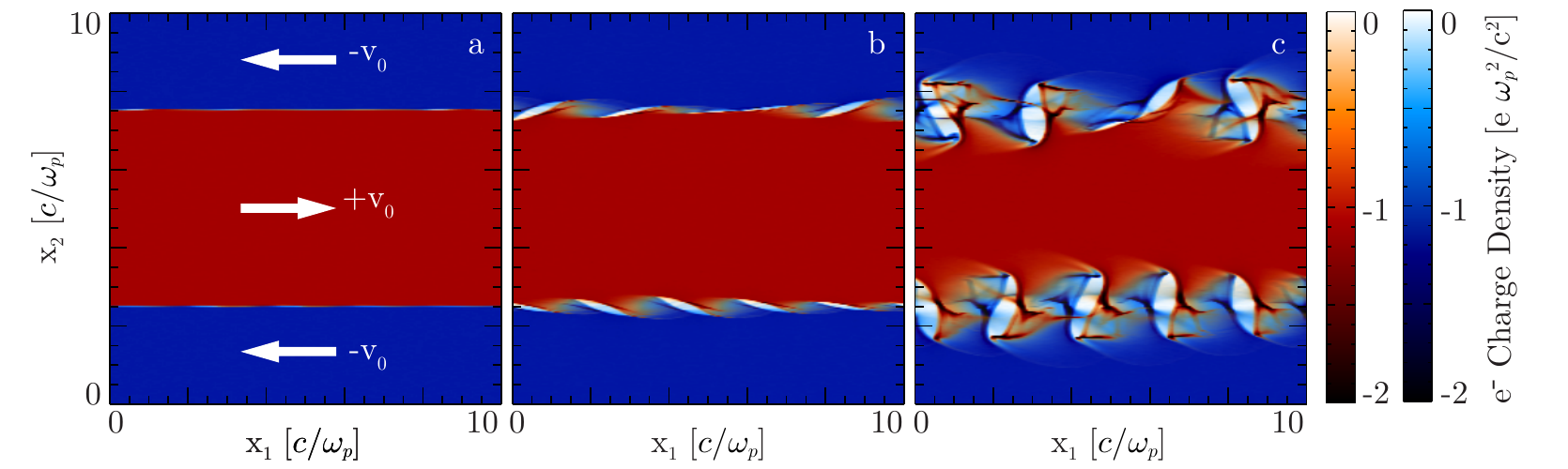}
\caption{Electron density structures at a) $\omega_pt = 35$, b) $\omega_pt = 45$, and c) $\omega_pt = 55$. The two flows stream with velocities $\mathbf{v}_0= \pm0.2c~\mathbf{e}_{x1}$}
\label{fig:r1_dens}
\end{figure}

\begin{figure}[t!]
\centering
\includegraphics[width=\columnwidth]{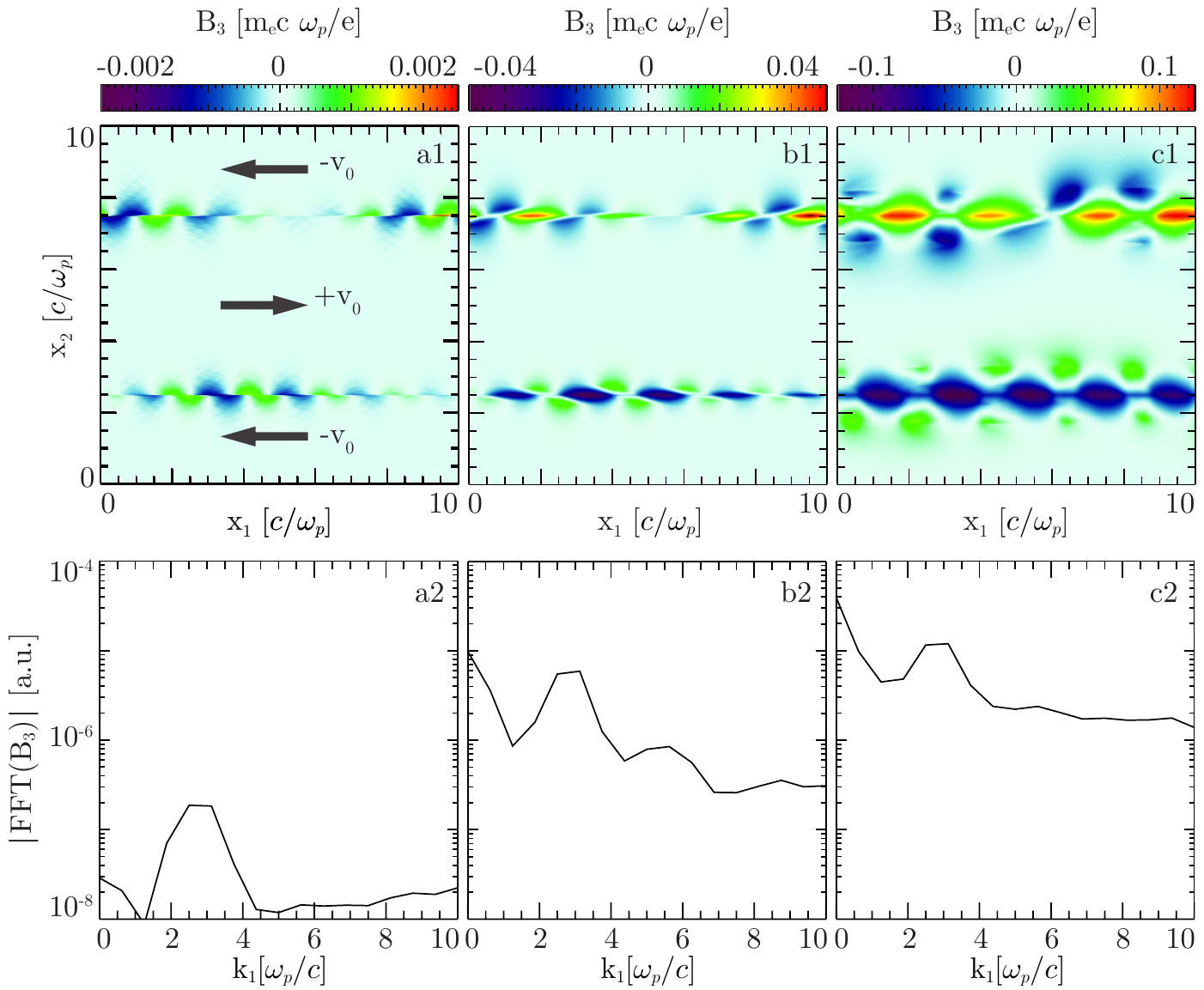}
\caption{(1) $B_3$ component of the magnetic field in the $xy$ plane and (2) corresponding average of the Fourier transform in $k_1$ at times a) $\omega_pt = 35$, b) $\omega_pt = 45$, and c) $\omega_pt = 55$.}
\label{fig:r1_b3}
\end{figure}

\begin{figure}[t!]
\centering
\includegraphics[width=\columnwidth]{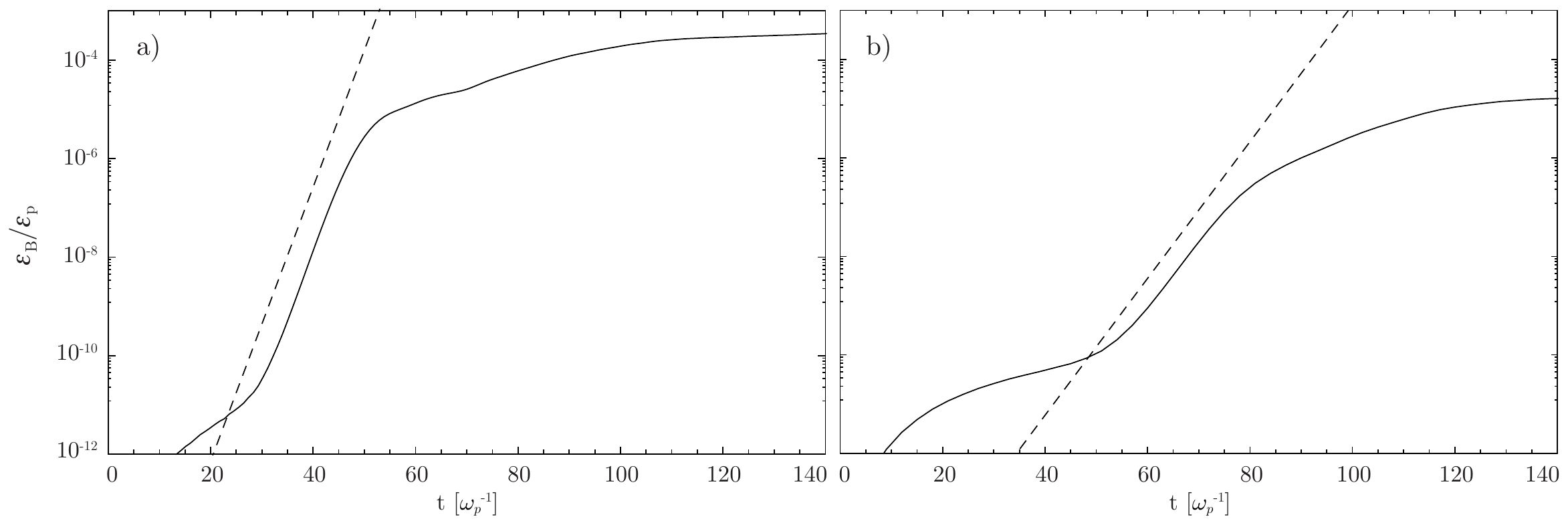}
\caption{Temporal evolution of the energy equipartition $\epsilon_B/\epsilon_p$ in scenarios a) shear between equal density flows, and b) shear between flows with density contrast $n_+/n_-=10$.}
\label{fig:b_evol}
\end{figure}

We begin by analyzing a subrelativistic shear scenario ($v_0=0.2~c$) where the counter streaming flows have equal densities. The evolution of the electron density of the system is depicted in Figure \ref{fig:r1_dens}, where the signature roll-up dynamics at the end of the linear phase of the KHI is observed. The protons of the system remain unperturbed (free-streaming) at these time scales due to their inertia. The wavelength of the growing perturbations in the electron density measure $2~c/\omega_p$, which corresponds to the wavelength of the fastest growing mode given by Eq.~(\ref{eq:kmax_gruzinov}). The magnetic field structure excited by the instability is shown in Figure \ref{fig:r1_b3}. The first inset of Figure \ref{fig:r1_b3} is taken during the linear phase of the KHI, showing the surface wave structure of the magnetic field, which is consistent with the two-fluid theory. The wavenumber parallel to the flow matches that of the theoretical fastest growing mode (Eq.~\ref{eq:kmax_gruzinov}), and the wave number perpendicular to the flow is evanescent. During the linear phase, the amplitude of the magnetic field grows exponentially (see Figure \ref{fig:b_evol} a) with a growth rate of $0.33 ~ \omega_p$, in close agreement with the theoretical prediction of Eq.~(\ref{eq:gr_gruzinov}) ($\Gamma = 0.35~\omega_p$). As the instability develops, the growing perturbations become strong enough to distort the sharp boundary between the shearing flows allowing them to mix. This mixing can no longer be treated with a fluid description, since the system dynamics becomes intrinsically kinetic. The signature of this kinetic regime is observed in Figure \ref{fig:r1_b3} b, where a dc component ($k_1 = 0$ mode) of the magnetic field begins to develop on top of the harmonic structure previously generated during the fluid regime. This dc magnetic field, is not unstable according to the fluid model as can be seen in Eq.~(\ref{eq:disp_rel_gruzinov}).
The evolution of dc magnetic field mode is clearly illustrated in Figure \ref{fig:r1_b3} (bottom inset), which shows the Fast Fourier Transform (FFT) spectrum in $k_1$ of the magnetic field in the system. At early times, the FFT spectrum reveals a peak around $k_1 = 3~\omega_p/c $, which corresponds to the unstable mode of the fluid regime (Figure \ref{fig:r1_b3} a)(bottom inset). At later times, however, when the fields previously developed during the fluid regime trigger the mixing/interpenetration between the two flows, the dc mode begins to develop (Figure \ref{fig:r1_b3} b)(bottom inset). Furthermore, when the instability saturates, the dc mode is the dominant component of the magnetic field, as shown in Figure \ref{fig:r1_b3} c (bottom inset). The physical picture underlying the growth and evolution of the dc mode of the magnetic field will be discussed later in Section \ref{sec:kinetic_regime}.

\subsubsection{Different density shear flows}

\begin{figure}[t!]
\centering
\includegraphics[width=\columnwidth]{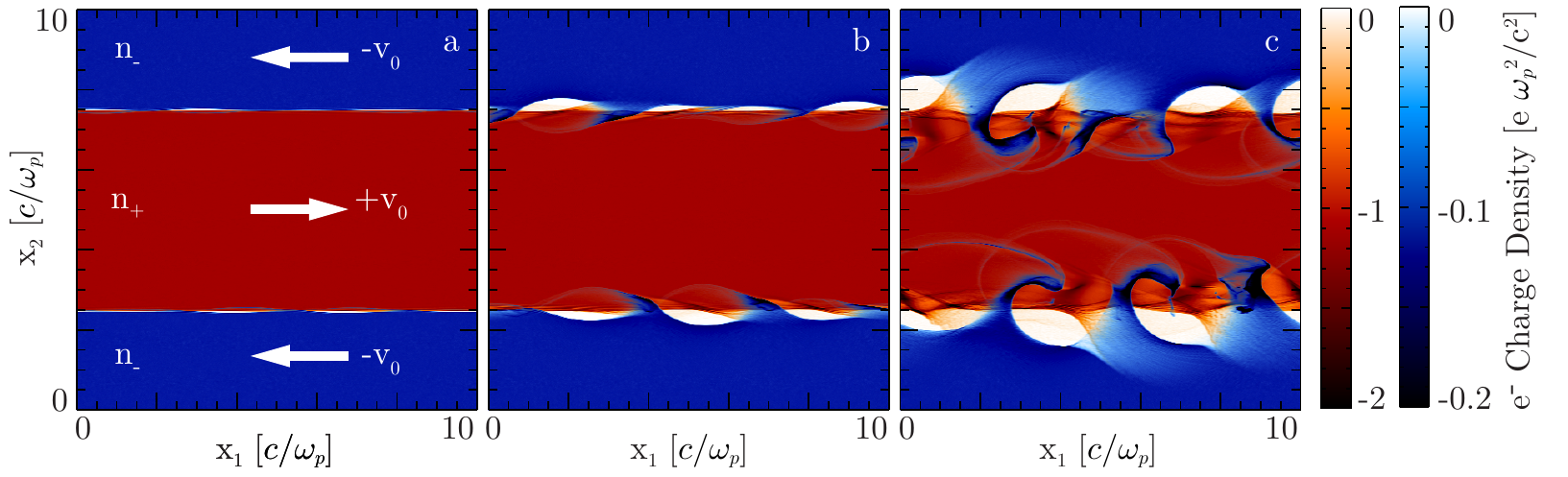}
\caption{Electron density structures for a shear flow with $n_{+}/n_{-} = 10$ at a) $\omega_pt = 60$, b) $\omega_pt = 75$, and c) $\omega_pt = 90$. The two flows stream with velocities $v_0= \pm0.2c$}
\label{fig:r10_dens}
\end{figure}

\begin{figure}[t!]
\centering
\includegraphics[width=\columnwidth]{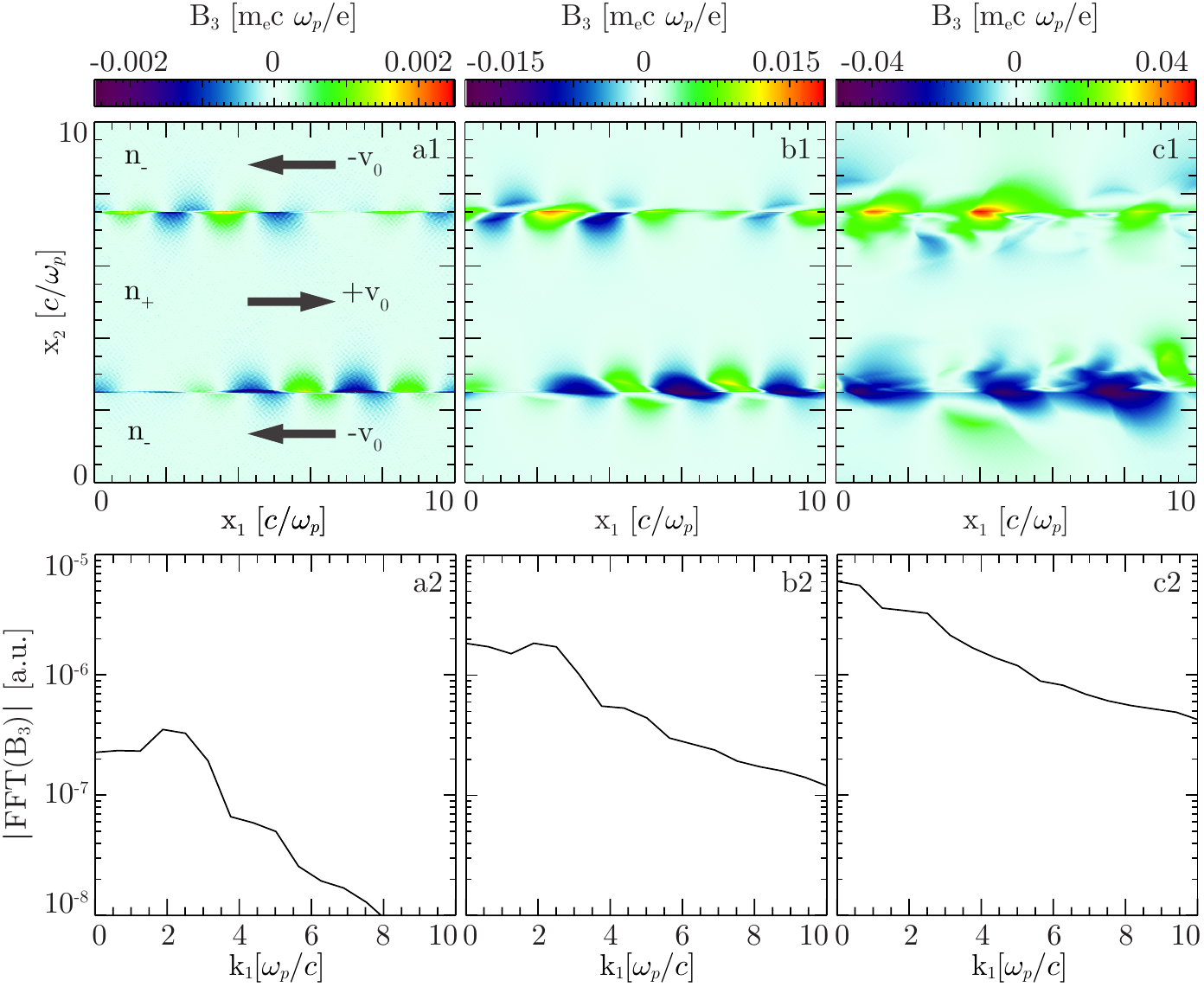}
\caption{(1) $B_3$ component of the magnetic field with $n_{+}/n_{-} = 10$ in the $xy$ plane and (2) corresponding average of the Fourier transform in $k_1$ at times a) $\omega_pt = 60$, b) $\omega_pt = 75$, and c) $\omega_pt = 90$.
}
\label{fig:r10_b3}
\end{figure}

The density contrast effects predicted by the theoretical two fluid model have also been verified with numerical simulations. Figure \ref{fig:r10_dens} shows the development of the electron density structures for a density contrast setting with $n_{+}/n_{-} = 10$. The KHI modulations that eventually turn into vortices are strongly manifested in the lower density plasma cloud (represented by the blue flow in Figure \ref{fig:stand_prop}). The typical length of these modulations is larger than those of the density symmetric case, as predicted by the theoretical model, measuring  $\lambda\simeq 3.3 c/\omega_{p+}$. This value agrees with the theoretical wavelength of the fastest growing mode, $\lambda_{\mathrm{max}}= 3.1 c/\omega_{p+}$. The self-generated magnetic field structure is represented in figure \ref{fig:r10_b3}, where the asymmetry in the evanescent behaviour of the surface mode in the different density regions can be observed. The growth rate of the instability is lowered with respect to the equal-density case and is in good agreement with the linear theory (figure \ref{fig:b_evol} (b)).

\subsection{Finite velocity shear gradient}

\begin{figure}[t!]
\centering
\includegraphics[width=0.6\columnwidth]{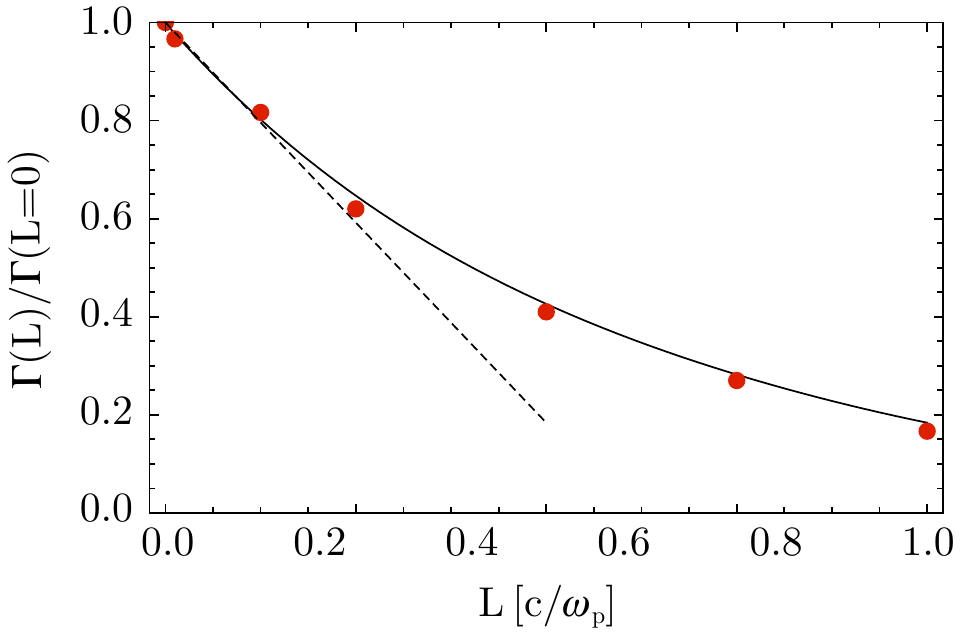}
\caption{Evolution of the maximum growth rate as a function of the gradient length. Dashed curve : expression (\ref{theoryGvsL}); plain curve : numerical algorithm; red dots : PIC simulations for $ v_0(x)/c = 0.2~tanh(x/L)$.}
\label{fig:gammavsL}
\end{figure}

The analytical treatment of the effect of a finite velocity shear gradient (smooth velocity shear profile) on the development of the electron-scale KHI is not trivial. The details of the underlying mathematics and numerics can be found in the Appendix.

We analyse the development of the electron-scale KHI for smooth velocity shear profile given by $v_0(x)=V_0 \tanh(x/L)$, where $L$ is the shear gradient length. An analytical solution for the growth rate of the instability can be found for small $k_\perp L$ (see Appendix), which reads
\begin{equation}
\label{theoryGvsL}
\frac{\Gamma_\mathrm{max}}{\Gamma_\mathrm{max}^0}\simeq 1-\frac{\sqrt{3}}{8}\pi k_{\perp}L,
\end{equation}
where $\Gamma_\mathrm{max}^0/\omega_p = \sqrt{1/8}$ is the growth rate obtained for a step velocity profile. The growth rate of the step velocity profile is recovered for $L=0$, and decreases for increasing $k_\perp L$. The wavenumber corresponding to the maximum growth rate follows a similar trend by slightly decreasing when the parameter $k_{\perp}L$ increases. For arbitrarily large $k_\perp L$, an exact numerical solution for the dispersion relation of the electron-scale KHI can be found, and we discuss a numerical scheme in the Appendix.

The above analytical and numerical results have been verified with PIC simulations. The setup of the simulations is identical to those previously described, only replacing the discontinuous velocity profile by the smooth function $v_0(x)=V_0 \tanh(x/L)$. This profile is also used in the numerical algorithm to solve the dispersion relation in the finite gradient shear scenario. The measurement of the maximum growth rate in the simulation is done by following in time the peak of the Fourier spectrum of one of the field structures during the linear phase of the instability. Figure \ref{fig:gammavsL} displays the maximum growth rate as a function of the gradient length for Eq. (\ref{theoryGvsL}), the numerical solution and the simulations results ($V_0 = 0.2~c$). For small values of the parameter $k_{\perp}L$, i.e., $L\ll 0.3~c/\omega_p$ ($k_{\perp}\sim 3~\omega_p/c$ for $V_0 = 0.2~c$), Eq. (\ref{theoryGvsL}) is in good agreement. For higher values of the gradient length, Eq. (\ref{theoryGvsL}) is not valid since it was derived in the first order ok $k_{\perp}L$. Nevertheless, the numerical solutions show a very good agreement with the simulations results, where we observe a decay of the maximum growth rate. Both the Eq. (\ref{theoryGvsL}) and the numerical solution have been verified with PIC simulations for various values of $V_0$ and are both in good agreement. We have observed the development of the electron-scale KHI with PIC simulations for $L$ up to $10~c/\omega_p$ \cite{grismayer13}.

\newpage
\section{dc magnetic field generation in unmagnetized shear flows}
 \label{sec:kinetic_regime}

\begin{figure}[]
\centering
\includegraphics[width=\columnwidth]{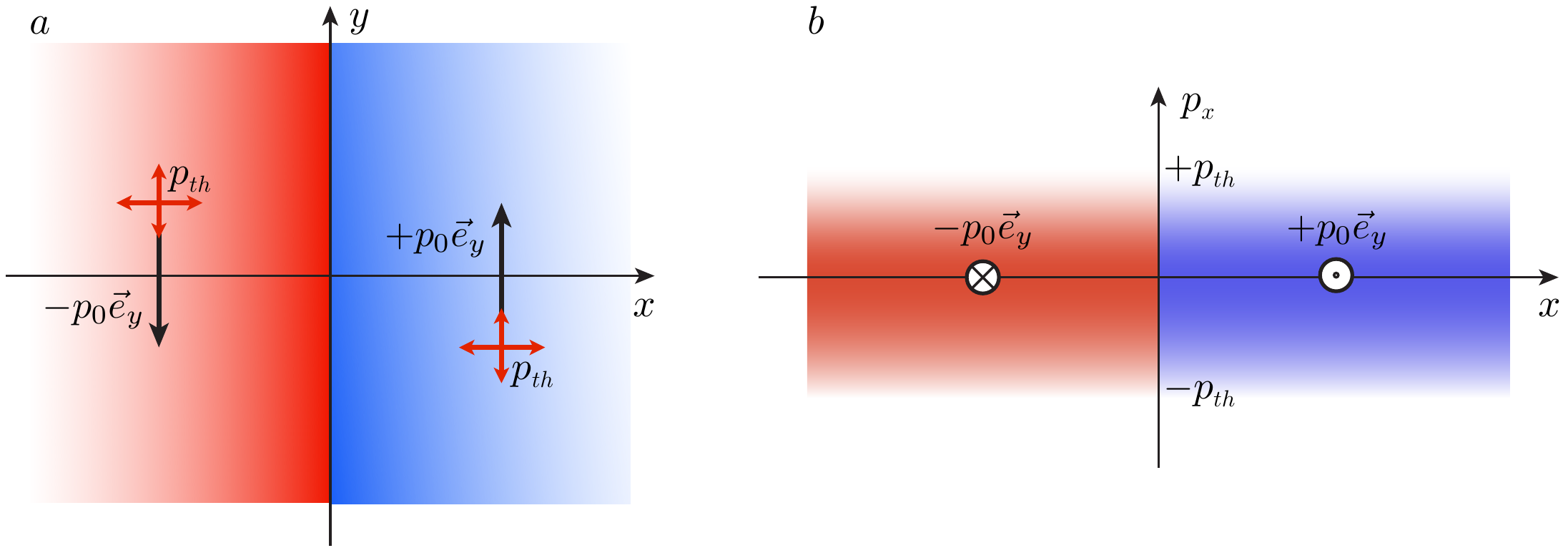}
\caption{a) Scheme of the initial condition b) Reduced geometry}
\label{fig:1dwarm}
\end{figure}

For the sake of completeness, we review in this Section the main results of the dc magnetic field generation mechanism in unmagnetised shear flows which are outlined in \cite{grismayer13}. We then present a more detailed analysis of the equipartition fields and the dependence of the equipartition number on the dimensionality of the model.

In the previous Section, numerical simulations showed the growth of a dc ($k=0$) magnetic field mode (Figure \ref{fig:r1_b3} (c) and Figure \ref{fig:r10_b3} (c)), which is not predicted by the linear fluid theory (Figure \ref{fig:unstable_modes}), $\Gamma'(k'=0))$ nor has it been previously identified in MHD simulations and only kinetic simulations \cite{alves12,grismayer13,boettcher} have been able to capture this mode. The growth of the dc magnetic field mode results from a current imbalance due to electron mixing across the shear interface, while the ion flows remain almost unperturbed due to their inertia. The orientation of the dc magnetic field peak is determined by the proton current structure. The mixing arises due to the deformation of the electron interface between the two flows which, in the linearized fluid calculations, is not accounted for and, in zeroth order, remains fixed. Alternatively, we find that the physics describing the formation of a dc mode can be modeled in a 1D reduced theory where an initial temperature drives the mixing effect.

%%%%%%%%%%%%%%%%%%%%%%%%%%%%%%%%%%%%%%%%%%%%%%%%%%%%%
\subsection{Warm shear flow}
\label{subsec:warm_scenario}
%%%%%%%%%%%%%%%%%%%%%%%%%%%%%%%%%%%%%%%%%%%%%%%%%%%%%

% theoretical model

We discuss here the temperature effect in a shear flow scenario, and its role in the generation of a dc magnetic field mode along the shear. For the sake of simplicity, and without loss of generality, we assume a simple sharp velocity shear transition between two plasmas with equal temperatures. We consider that the temperature is sufficiently high such that the electron thermal expansion time scale is much faster than the electron expansion induced by the onset of the fluid KHI in an equivalent cold scenario. The theoretical setting of the system is illustrated in Figure \ref{fig:1dwarm}-a. We consider only the electron thermal velocity, neglecting the thermal velocity of the protons due to their inertia. Since we are interested in describing the dc phenomena, all derivatives along the $x$ direction vanish, reducing the system to the 1D problem displayed in Fig. \ref{fig:1dwarm}-b. We, therefore, consider the purely one-dimensional case where the particles can move along $x$, as in Figure \ref{fig:1dwarm}-b. Initially all the fields are zero and we assume a warm initial plasma with a tangential shear flow identical to the one described in Sec.\ref{sec:lin_develop_khi} with an initial temperature such as $v_{th}\ll v_0$. This setting is not in Vlasov equilibrium and it is clear that the thermal expansion of the electrons across the shear surface (ions are assumed to be cold and free streaming) leads to an imbalance of the current neutrality around the shear surface, forming a dc magnetic field in z direction.  The initial corresponding electron distribution function reads
\begin{equation}
f(x,v_x,v_y,v_z,t=0)=f_0(v_x,v_y-v_0\mathrm{sign}(x),v_z) 
\end{equation}
The situation can be seen as two thermal plasmas with shearing counter propagating fluid velocities. The thermal expansion of the electrons (the ions, due to their inertia, are assumed free streaming) across the shear will transport an electron current on the order of $en_0v_0$, with a characteristic width of $v_{thx}t$. This should lead, at early times, to the formation of a field $B_z$ around the shear of width $v_{thx}t$ and magnitude of $\mu_0en_0v_0v_{thx}t$. This is the underlying physical picture of the dc magnetic field growth. Due to the dimensionality of the problem, it is clear that $E_z, B_x, B_y$ remain zero. The reduced set of equation is 

\begin{eqnarray}
\label{maxFar1d}
-\frac{\partial B_z}{\partial t} &=& \frac{\partial E_y}{\partial x} \\
\label{maxamp1d}
-\frac{\partial B_z}{\partial x} &=& \mu_0 J_y+\frac{\partial E_y}{\partial t}\frac{1}{c^2}\\
\label{maxamp1d2}
\mu_0 J_x&=&-\frac{\partial E_x}{\partial t}\frac{1}{c^2}
\end{eqnarray}

\begin{equation}
\label{vlasov}
\frac{\partial F}{\partial t} + v_x\frac{\partial F}{\partial x}-\frac{e}{m}(\mathbf{E}+\mathbf{v}\times\mathbf{B_z}).\frac{\partial F}{\partial \mathbf{v}}=0
\end{equation}
where
\begin{equation}
F(x,v_x,v_y,t)=\int dv_z f(x,v_x,v_y,v_z,t)
\end{equation}
The formal solution of the Vlasov equation Eq. (\ref{vlasov}) is
\begin{equation}
F(x,v_x,v_y,t)=F_0(x_0,v_{x0},v_{y0})
\end{equation}
where $x_0$, $v_{x0}$ and $v_{y0}$ denote the position and velocities of an electron at $t=0$ and $f_0=\int dv_{z0}F_0$. At early times, if we assume that the induced fields are sufficiently small that we can neglect the change of momentum of the electrons, the distribution can be solved along the free streaming orbits, i.e., $x=x_0+v_{x0}t, v_{x}=v_{x0}, v_{y}=v_{y0}$. For the sake of simplicity, we divide the initial electron distribution in two parts, $F_0=F_0^{-}(x_0<0)+F_0^{+}(x_0>0)$, corresponding to the two initially separated flows. In the approximation of free streaming orbits, the electron currents read
\begin{eqnarray}
J_{e,y}^{\pm}\simeq-e\int dv_{y}v_{y}\int dv_{x} f_0^{\pm}(x-v_xt,v_x,v_y\mp v_0).
\end{eqnarray}
With $f_0^{\pm}(x_0,v_{x0},v_{y0}\mp v_0)=n_0f_M(v_{x0})f_M(v_{y0}\mp v_0)$, where $f_M(v)=e^{-v^2/2v_{th}^2}/
\sqrt{2\pi} v_{th}$ represents the Maxwellian velocity distribution, we obtain for the electron currents
\begin{eqnarray}
J_{e,y}^{\pm}&\simeq& \mp en_0v_0\int_{\mp x/t}^{\infty} dv_{x} f_M(v_x) \\
&\simeq& \mp \frac{ev_0n_0}{2}~\mathrm{erfc}\left(\frac{\mp x}{\sqrt2v_{thx}t}\right)
\end{eqnarray}
The total current is obtained by adding the unperturbed proton currents, this yields 
\begin{eqnarray}
J_y&=&ev_0n_0\left[-2+\mathrm{erfc}\left(\frac{x}{\sqrt2v_{thx}t}\right)\right]~~~~~x\le 0 \\
J_y&=&ev_0n_0~\mathrm{erfc}\left(\frac{x}{\sqrt2v_{thx}t}\right)~~~~~x> 0 
\end{eqnarray}
The magnetic field is then simply given by the Maxwell-Ampere equation, Eq. (\ref{maxamp1d}), where the displacement current is neglected
\begin{eqnarray}
B_z&\simeq& -\mu_0\int dx J_y \\
&\simeq& -e\mu_0v_0n_0\sqrt{2}v_{thx}t\left[\xi\mathrm{erfc}(\xi)-\frac{e^{-\xi^2}}{\sqrt{\pi}}\right]~~ ;\xi=\frac{|x|}{\sqrt{2}v_{thx}t}
\label{eq:bdc}
\end{eqnarray}
We verify then that the thermal expansion, transporting currents across the shear, induces a magnetic field that grows linearly with time. Its typical width, on the order  $\sqrt{2}v_{thx}t$ and its peak $B_z(x=0)=\sqrt{2/\pi}\mu_0en_0v_0v_{thx}t$, is in agreement with our previous estimates. The associated magnetic energy growing in the system is given by 
\begin{eqnarray}
\epsilon_B = \int dx \frac{B_z^2}{2\mu_0} \simeq 0.156\sqrt{2}\mu_0(en_0v_0)^2(v_{thx}t)^3,
\label{eq:warm_bene}
\end{eqnarray}
with $\int du \left[ |u| \mathrm{erfc}(|u|)-e^{-u^2}/\sqrt{\pi}\right]^2\simeq0.156$.

\begin{figure}[]
\centering
\includegraphics[width=0.9\columnwidth]{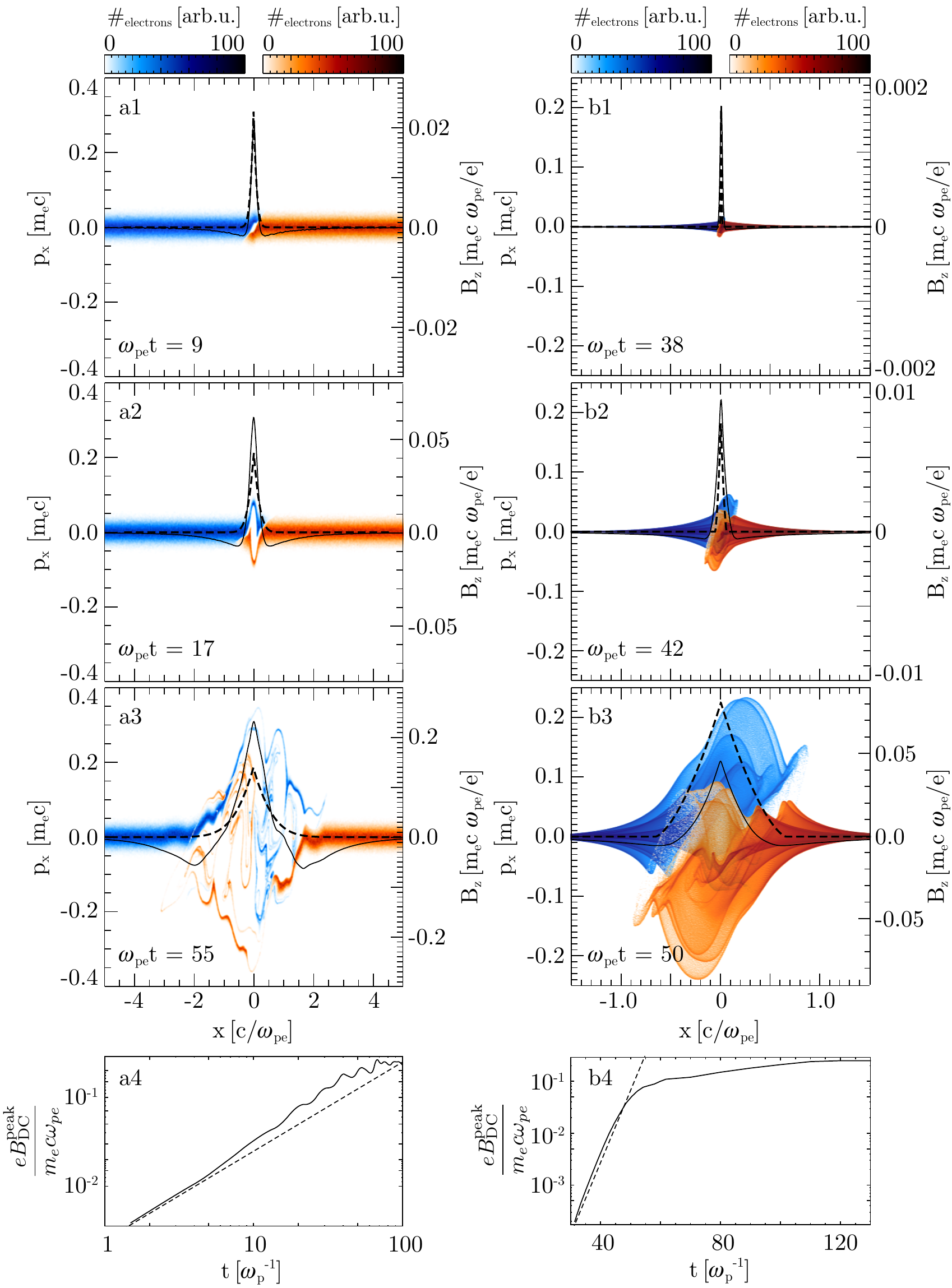}
\caption{Evolution of the electron phasespace (insets a1-3 and b1-3) and dc magnetic field peak (insets a4 and b4); log-log is used to display linear dc peak evolution in inset a4, and log-linear is used to display exponential evolution in inset b4. Left: 1D warm shear flow with $v_0 = 0.2c$ and $v_{th} = 0.016c$. Right: 2D cold shear flow for the same $v_0$. The blue (red) color represents the electrons with a negative (positive) drift velocity $v_0$. The self-consistent dc magnetic field is represented by the solid curve, whereas the dashed curve represents the magnetic field given by the theoretical model.}
\label{fig:p1x1_b3_small_vth}
\end{figure}

% Feedback of the fields over the plasma expansion
As was pointed out, this derivation is only valid as long as the orbits of the electrons do not diverge much from the free streaming orbits, i.e., as long as the electric and magnetic fields that develop in the expansion process do not affect the free motion of the particles. In fact, there are two phenomena that affect the growth of the magnetic field. First, the electrons will eventually feel the induced magnetic field which tends to push more electrons across the shear via the $\mathbf{v}_0\times\mathbf{B}_{z}$ force. This will increase the rate of current transported and, thus, will increase the growth rate of the magnetic field. One should note that, at first, only the magnetic field acts on the electrons since the electric field $E_x$ remains zero in our model: the initial temperature is uniform in space and therefore, there are as many electrons crossing from the left than from the right. The charge neutrality is then conserved in the system. A crude estimate of the time at which our model breaks can be made considering that only the slow electrons, initially around the shear, will experience a strong velocity change due to the peaked shape of the magnetic field. Therefore, the model will break down approximatively when an electron initially at rest (around $x=0$) acquires a velocity change on the order of $v_{thx}$, which corresponds to a strong distortion of the Maxwellian distribution function around the shear. This can be written for the velocity change as
\begin{equation}
\label{eq:warm_model_break_down}
|\delta v_x| \sim \frac{ev_0}{m}\int_0^t dt' B_z(0,t') \sim v_{thx}\left(\frac{v_0}{c}\right)^2\frac{(\omega_pt)^2}{\sqrt{2\pi}}.
\end{equation}
It follows that the model is valid until $\omega_p t \sim (2\pi)^{1/4}c/v_0$. Second, the growth of the magnetic field induces an electric field $E_y$  through the Maxwell-Faraday equation Eq. (\ref{maxFar1d}). We can estimate through the characteristic time $t$ and length of the problem $v_{thx}t$, the magnitude of the field,  $E_y\sim v_{thx}B_z$. Inserting the value of $E_y$ into the Maxwell-Ampere equation, we find the displacement current term $\partial E_y/\partial t$ leads to a $(v_{thx}/c)^2$ correction to the dc magnetic field peak at early times. However, the displacement current tends to increase the electron current on either sides of the shear interface, eventually building the dc magnetic field side wings observed at later times (Fig. \ref{fig:p1x1_b3_small_vth} a3-b3).

In order to verify our analytical calculations and to further investigate the phase at which the electrons deviate from their free streaming orbits, we have simulated a shear flow between warm electron-proton plasma slabs with a realistic mass ratio $m_p/m_e=1836$ until the dc magnetic field structure saturates on the electron time scale. The Debye length is resolved in the 1D simulations ($\Delta x = \lambda_D$) and we have used 1000 particles per cell. These PIC results have been presented in \cite{grismayer13} but are reproduced here for sake of completeness. Fig. \ref{fig:p1x1_b3_small_vth} (a1-a3) shows the time evolution of the $x p_x$ phase space and the magnetic field for $v_{th}=0.016 c$ and $v_0=0.2c$. At earlier times $\omega_{pe}t=9$ (Fig. \ref{fig:p1x1_b3_small_vth} (a1)) an excellent agreement between the model and the simulation is observed. According to Eq. \ref{eq:warm_model_break_down}, the model breaks down for times larger than $\omega_{pe}t  10$. The deviation from the Maxwell equilibrium of the distribution function in the shear region is shown at $\omega_{pe} t=17$. The model underestimates the magnitude of the magnetic field and one can clearly observe the distortion of the distribution function in the field region. As the magnetic field grows, the Larmor radius ($r_L$) of the electrons crossing the shear interface decreases. When the minimum $r_\mathrm{L,min}$ (associated to the peak of $B_\mathrm{DC}$) becomes smaller than the characteristic width of the magnetic field $l_\mathrm{DC}$, the bulk of the electrons becomes trapped by the magnetic field structure. This is illustrated in Fig. \ref{fig:p1x1_b3_small_vth} (a3) at $\omega_{pe} t= 55$. The magnetic trapping prevents the electron bulk expansion across the shear (that drives the growth of the magnetic field), saturating the magnetic field. An estimate of the saturation can be obtained by equating $r_\mathrm{L,min} \sim l_\mathrm{DC}$. From Eq.~(\ref{eq:bdc}), it is possible to write the magnetic field as $B_\mathrm{DC}(x,t)=4\pi en_0\beta_0w(x,t)$, where $w(0,t)$ should be interpreted as the characteristic width of the field. With $l_\mathrm{DC}\sim w(0,t)$, $r_\mathrm{L,min}=mv_0\gamma_0/eB_\mathrm{DC}(0,t)$, we find that $l_\mathrm{DC}\sim c\sqrt{\gamma_0}/\omega_{pe}$ giving the saturation level of the magnetic field as 
\begin{equation}
\label{eq:dcsaturation}
\frac{eB_\mathrm{DC}^\mathrm{sat}}{m_ec\omega_{pe}}\sim \beta_0\sqrt{\gamma_0}.
\end{equation}
This scaling has been verified for 1D simulations (see Fig. 3 in \cite{grismayer13}) for which the best fit function matching the simulation is $eB_\mathrm{DC}^\mathrm{sat}/m_ec\omega_{pe}= 1.9\beta_0\sqrt{\gamma_0}$.

%%%%%%%%%%%%%%%%%%%%%%%%%%%%%%%%%%%%%%%%%%%%%%%%%%%%%%
%\subsection{2D cold shear flow}
%\label{subsec:cold_scenario}
%%%%%%%%%%%%%%%%%%%%%%%%%%%%%%%%%%%%%%%%%%%%%%%%%%%%%%

\subsection{Cold shear flow}

In the absence of an initial temperature, an alternative mechanism is needed to drive the electron mixing across the shear surface that, in turn, generates the dc field. This mechanism is the cold fluid KHI that has been thoroughly discussed in Sec. \ref{sec:fluid_regime}. In fact, in the warm shear flow scenario, both the cold fluid KHI and the electron thermal expansion can contribute to the generation of the dc field. This occurs when the typical length of the dc field due to the thermal expansion ($l_\mathrm{DC}$) after a few e-foldings of the cold fluid KHI ($T_\mathrm{KHI-growth} = N_\mathrm{e-foldings}/\Gamma_\mathrm{max}$, where $N_\mathrm{e-foldings}$ is on the order of 10) is on the order of the relativistic electron skin depth, i.e., $v_{th} T_\mathrm{KHI-growth} \sim \sqrt{\gamma_0}c/\omega_{pe}$. Therefore, the cold fluid KHI dominates the electron mixing in the limit  
\begin{equation}
v_{th} T_\mathrm{KHI-growth} \ll \sqrt{\gamma_0}\frac{c}{\omega_{pe}}.
\end{equation}
 
For a two dimensional cold plasma undergoing the KHI, the electron distribution function can be written as 
\begin{equation}
f(x,y,v_x,v_y,v_z,t)=n_0\delta(v_x-v_{xfl}(x,y,t))\delta(v_y-v_{yfl}(x,y,t))\delta(v_z)
\end{equation}
where $v_{xfl},v_{yfl}$ correspond to the velocity field solutions of the fluid theory. 
%The electron orbits in the KHI field structure reveal that electrons will eventually cross the shear surface, unbalancing the current neutrality. 
In this case, the self-generated KHI fields play the role of an effective temperature that transports the electrons across the shear surface, while the protons remain unperturbed, inducing a dc component in the current density, and hence in the fields. We then have to solve the evolution of the distribution function and show that the current density $J_y$, averaged over a wavelength $\lambda=2\pi/k_\parallel$, has a non zero dc part. We follow the same approach as before and calculate the average distribution function defined as:
\begin{equation}
\label{eq:fave}
F(x,v_x,t)=\frac{1}{\lambda}\int dv_y\int dv_z \int_{\lambda} dy f(x,y,v_x,v_y,v_z,t).
\end{equation}
To obtain analytical results we will assume that the linearly perturbed fluid quantities are purely monochromatic, which is equivalent to assume that after a few e-foldings, the mode corresponding to $k_\parallel=k_{\parallel \mathrm{max}}$ dominates with a growth rate of $\Gamma = \Gamma_{max}$. We then write $v_{yfl}\simeq v_0(x)$ and $v_{xfl}=\bar{v}_{xfl}\sin(k_\parallel y)e^{-k_\perp|x|+\Gamma t}$, where $\bar{v}_{xfl}$, the amplitude of the velocity perturbations at $t=0$, is associated to the small thermal fluctuations (small enough to ensure that the thermal expansion is negligible over $T_\mathrm{KHI-growth}$). Inserting $v_{xfl},v_{yfl}$ into Eq.~(\ref{eq:fave}), we obtain
\begin{eqnarray}
F(x,v_x,t)=\frac{n_0}{\pi v_{max}\sqrt{1-\xi^2}},
\end{eqnarray}
where $\xi(x,v_x,t)=v_x/v_{max}(x,t)$ with $v_{max}(x,t)=\bar{v}_{xfl}e^{-k_{\perp}|x|+\Gamma t}$.
We observe that the development of 2D cold KHI reveals close similarities with the 1D hot model previously described. In the 2D KHI, averaging the distribution in the direction of the flow shows that the perturbation gives rise to a spread in $v_x$ that may be interpreted as an effective temperature. The spread in $v_x$ decays exponentially away from the shear and grows exponentially with time. The mean velocity is zero and the effective temperature associated to this distribution function is defined as 
\begin{equation}
V_{eff}^2(x,t)=\frac{1}{n_0}\int dv_x v_x^2 F(x,v_x,t)=\frac{v_{max}^2}{2}. 
\end{equation}
One can then expect a similar physical picture as in the hot shear scenario and, as a result, the emergence of dc components in the fields which are induced by the development of the unstable KHI perturbations. The evolution of the phase space in Fig. \ref{fig:p1x1_b3_small_vth} illustrates the similarity between the warm 1D (insets a1-3) and cold 2D (insets b1-3) scenarios. 

\begin{figure}[h!]
\centering
\includegraphics[width=0.7\columnwidth]{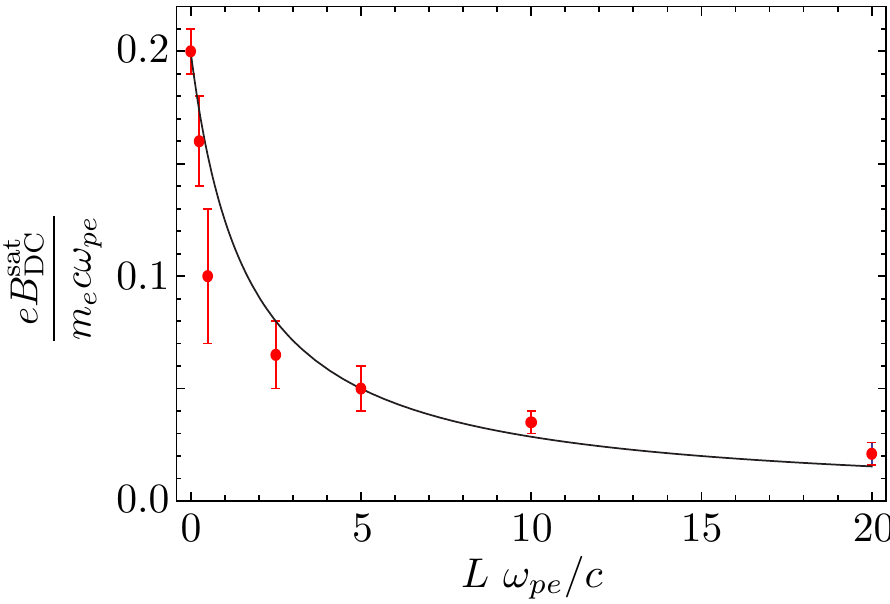}
\caption{\label{Bdcsatsmooth} Magnitude of the dc magnetic field peak at saturation as a function of the initial shear gradient length for $v_0/c=0.2$. The error bars are associated to the fluctuations of the peak value in the saturation stage. The dash-line represents the best fit curve to the simulation results, given by $eB_\mathrm{DC}^\mathrm{sat}/m_e c \omega_{pe}=0.2/(1+0.6L\omega_{pe}/c)$ }
\end{figure}

The challenge in this scenario is to determine how such a distribution function expands across the shear surface due to the complexity of the orbits in the fields structure (multidimensional fields with discontinuities at $x=0$). One can however overcome this difficulty by solving the expansion along approximate orbits. This procedure, although not self-consistent, gives rich qualitative and quantitative insight regarding the features of the current that develops around the shear interface. In the $x p_x$ phase space, the electrons describe outward-spiraling growing orbits, since they are drifting across the standing growing perturbation. In the region where the electron mixing occurs, we assume electron orbits given by $x\sim x_0+(v_{x0}/\Gamma) e^{\Gamma t}$ and $v_x\sim v_{x0}e^{\Gamma t}$ where $x_0$ and $v_{x0}$ are the position and velocity of a particle at the time $t_0$ when the instability begins. 
\begin{eqnarray}
J_{e,y}^{\pm}(x,t)&=&-\frac{e}{\lambda}\int_{\lambda} dy\int  dv_x\int dv_y v_y  f^{\pm}(x,y,v_x,v_y,t) \\
\label{eq:jdc_int}
J_{e,y}^{\pm}(x,t)&\simeq&\mp ev_0\int_{\mp x\Gamma}^{v_{max}^0} dv_x F(x,v_x,t) \\
&\simeq&ev_0n_0\left[\frac{1}{2}\pm\frac{1}{\pi}\arcsin\left(\frac{x\Gamma}{v_\mathrm{max}^0}\right)\right]
\end{eqnarray} 
where $x\Gamma \in [-v_\mathrm{max}^0,v_\mathrm{max}^0]$ and $v_\mathrm{max}^0(t)=v_\mathrm{max}(x=0,t)$ that represents the maximum velocity of a particle that was originally in the vicinity of the shear. The limits of the integral in Eq. (\ref{eq:jdc_int}) represent the deformation of the boundary between the two flows on a characteristic distance of $v_\mathrm{max}^0/\Gamma$ as the instability develops. In the fluid theory, the boundary remains fixed, precluding the development of the dc mode. We then find the total current density by summing the proton contribution and integrating to obtain the induced dc magnetic field:
\begin{eqnarray}
\label{bdc_cold_expr}
B_\mathrm{DC}(\pm x \ge 0,t)=\mp8en_0\beta_0x\left[\arcsin(\zeta)\mp\frac{\pi}{2}\pm\sqrt{\frac{1}{\zeta^2}-1}\right]
\end{eqnarray}
with $\zeta=\Gamma x/v_\mathrm{max}^0$. The peak of the dc magnetic field is located at $x=0$ where the expression above reduces to $B_\mathrm{DC}(0,t)=8e\beta_0n_0v_\mathrm{max}^0(t)/(\pi\Gamma)$ and thus grows at the same rate as the KHI fields. One can verify in Fig. \ref{fig:p1x1_b3_small_vth} (b1-b2) that Eq.~(\ref{bdc_cold_expr}) shows reasonable agreement with the 2D simulations. This derivation neglects the dc Lorentz force on the electron trajectories, which makes this model valid as long as the induced dc fields remains small compared to the fluid fields associated to the mode $k_{\parallel \mathrm{max}}$. The peak of the $B_\mathrm{DC}$ field is proportional to $v_\mathrm{max}^0(t)$ that represents the maximum value of the fluid velocity $v_{xfl}$, which obeys $v_{xfl}=-ie(E_{xfl}+\beta_0B_{zfl})/m_e\gamma_0(\omega-k_\parallel v_0)$. From the linear fluid theory, one can compute the ratio $B_{zfl}/E_{xfl}$ from which we deduce that for sub-relativistic shears ($\gamma_0 \sim 1$), $|v_{xfl}|\sim ec\sqrt{2/7}|B_{zfl}|/m_e\omega_{pe}v_0$ implying $B_\mathrm{DC}\sim(8/\sqrt{7}\pi) B_{zfl}$ and that for ultra-relativistic shears ($\gamma_0 \gg 1$), $|v_{xfl}|\sim e|B_{zfl}|/m_ec\sqrt{2}\omega_{pe}\gamma_0^{3/2}$ yielding $B_\mathrm{DC}\sim(4/\pi) B_{zfl}$. 
We conclude that the induced dc magnetic field is always on the same order as the fluid fields and thus its consequences to KHI development cannot be neglected. As the dc field evolves, electrons start to get trapped and we expect a level of saturation similar to the saturating level obtained in the 1D model. This has been verified by the simulations. The comparisons between the saturation level of the 1D, 2D and 3D simulations are shown in Fig. 3 in \cite{grismayer13}, also verifying the $\beta_0\sqrt{\gamma_0}$ scaling. One can also compute the equipartition number related to the magnetic field at saturation. Using Eq.~(\ref{eq:dcsaturation}) one finds, 
\begin{eqnarray}
\epsilon =\frac{\int_{l_\mathrm{DC}} dx {B_\mathrm{DC}^\mathrm{sat}}^2/8\pi}{\int_{l_\mathrm{DC}} dx n_0(m_e+m_p)c^2(\gamma_0-1)} \sim \frac{1}{2}\frac{m_e}{m_p}\frac{\gamma_0+1}{\gamma_0} 
\end{eqnarray}
which is similar to the equipartition number found for the Weibel instability \cite{medvedev99}. Our derivation for the saturation of the dc magnetic field allows also to recover the empirical estimate of \cite{alves12} already given for such a shear scenario. When a smooth shear is considered, the electron KH still develops as we have shown in Sec. \ref{sec:fluid_regime}. We verified that the initial electron transport across the shear, due the development of the instability, is the mechanism triggering the magnetic field generation, therefore validating the physics captured by our model. At saturation the dc magnetic field has a typical width on the order of the initial shear gradient length. Keeping the same arguments we have used to derive the approximate value of the dc field at saturation, i.e., $r_{\mathrm{L,min}}\sim L$ implies
\begin{equation}
\label{eq:dcsaturationsmooth}
\frac{eB_\mathrm{DC}^\mathrm{sat}}{m_ec\omega_{pe}}\sim \frac{B_\mathrm{DC}^\mathrm{sat}(\tilde{L}=0)}{\tilde{L}}
\end{equation}
with $\tilde{L}=L\omega_{pe}/c\sqrt{\gamma_0}$. Such a scaling has been verified for $L\omega_{pe}/c\gg 1$ and the comparison between the crude estimate and the simulations is presented in Fig. \ref{Bdcsatsmooth}. Interestingly, the dc magnetic field remains stable beyond the electron time scale and persists up to 100s $\omega_{pi}^{-1}$ (see Fig. 1 in \cite{grismayer13}). Eventually the protons will drift away from the shear surface due to the magnetic pressure, broadening the dc magnetic field structure and lowering its magnitude.

\newpage
\section{Particle acceleration}
\label{sec:energy}

We investigate in this Section the acceleration of particles due to the development of electron-scale shear instabilities. Particle acceleration in shear flows has been previously investigated by many authors, mainly related to astrophysical scenarios \cite{Berezhko,Jopikii,Webb,Ostrowski,Rieger,Rieger2}. The shear acceleration mechanism (\cite{Rieger2}) is based on the idea that energetic particles may gain energy by systematically scattering off of moving small-scale magnetic field irregularities. These irregularities are thought to be embedded in a collisionless shear flow such that their velocities correspond to the local flow velocity. In the presence of velocity shear, the momentum of a particle travelling across the shear changes and the acceleration process essentially draws on the kinetic energy of the background flow. 
In the shear flows we discussed in the previous sections, fluid and kinetic effects lead to the emergence of organized electric and magnetic fields that are maintained in the shear region up to ion time scale. The electrons flowing in the shear region experience strong acceleration in these fields and also emit strong radiation while gyrating in the dc magnetic field. This acceleration process therefore differs from the shear acceleration mechanism of \cite{Rieger2}.

In order to investigate the acceleration of electrons in the shear due to the self-generated fields, we performed a 2D simulation of a relativistic cold shear flow, $\gamma_0=3$, $v_{th}/c=10^{-3}$. The simulation domain dimensions are $250\times 2000 (c/\omega_p)^2$, resolved with 10 cells per electron skin depth ($\Delta x_1=\Delta x_2 = 0.1 c/\omega_p$) and 36 particles per cell per species are used. The shear flow initial condition is set by a velocity with $+p_0\vec{e_1}$ for $x_2>0$ and a symmetric flow with $-p_0\vec{e_1}$ for $x_2<0$. We impose periodic and absorbing boundary conditions in the $x_1$ and $x_2$ directions, respectively. The transverse direction $x_2$ has been extended up to $2000 c/\omega_p$ with absorbing boundary condition, in order to avoid particle recirculation over the shear region which tends to produce unphysical additional acceleration. The dimension of the simulation box allows us then to follow the evolution of the system until $\omega_p t\sim 1000$, time at which some particles approach the boundaries of the box. 

The growth rate and fast growing mode in the early development of the relativistic electron-scale KHI was found to agree with the theoretical predictions of Sec. \ref{sec:fluid_regime}. As explained in Sec. \ref{sec:kinetic_regime}, the nonlinear development of the instability drives the mixing between electron populations at the shear interface and generates dc components in the fields. At full saturation of the instability, the persistent electric and magnetic field components are on the order of $\sqrt{\gamma_0}~m_e\omega_pc/e$. During the early stage of the instability the oscillating fields are responsible for acceleration and deceleration of the electrons. This results in a slight temperature increase of the plasma and the electron distribution function widens around the mean energy $\gamma_0$. Once the instability saturates, the electrons can experience strong acceleration due to the long-lived electric field structures in the shear region. Furthermore, the dc magnetic field, that only remains intense in the shear region, provides one of the mechanisms to curve the electron trajectories and hence takes part in the thermalization and isotropization of the electron distribution function.
\begin{figure}[]
\centering
	\includegraphics[width=0.7\columnwidth]{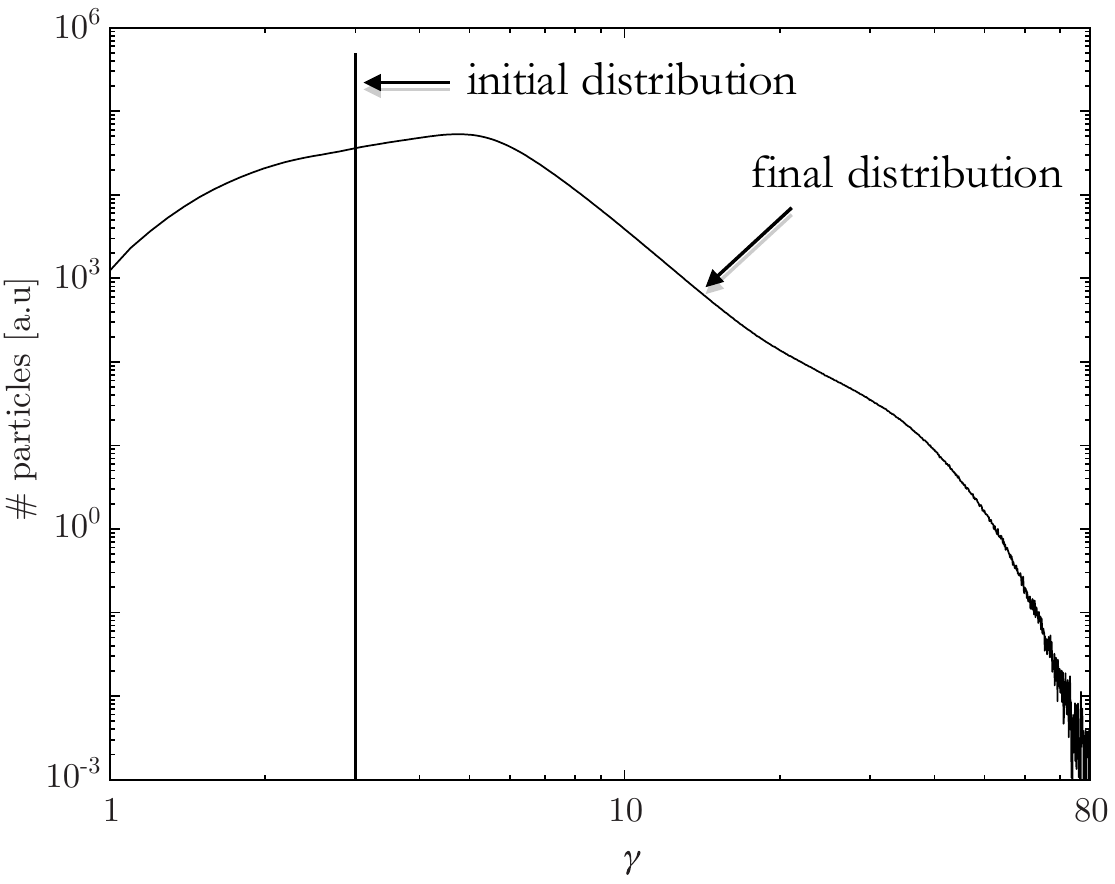}
	\caption{Electron energy distribution function at $\omega_pt=0$ and $\omega_pt=1000$.}
	\label{fig:gamdist}
\end{figure}
The energy distribution function of the electrons is plotted in Fig. \ref{fig:gamdist} at at $\omega_pt=0$ and $\omega_pt=1000$. The final distribution can be separated in three distinct parts. The low energy part, $1<\gamma< 5$, corresponds to a thermal Juttner distribution, $f(\gamma)\sim\gamma^2e^{-\mu\gamma}$ with $\mu\sim 1/\gamma_0$. The medium energy part exhibits a power law $\gamma^{-5}$ up to $\gamma \sim 25$. Finally after the elbow of the distribution one notices a hot tail that extends up to $\gamma = 80$.

\begin{figure}[]
\centering
\includegraphics[width=0.7\columnwidth]{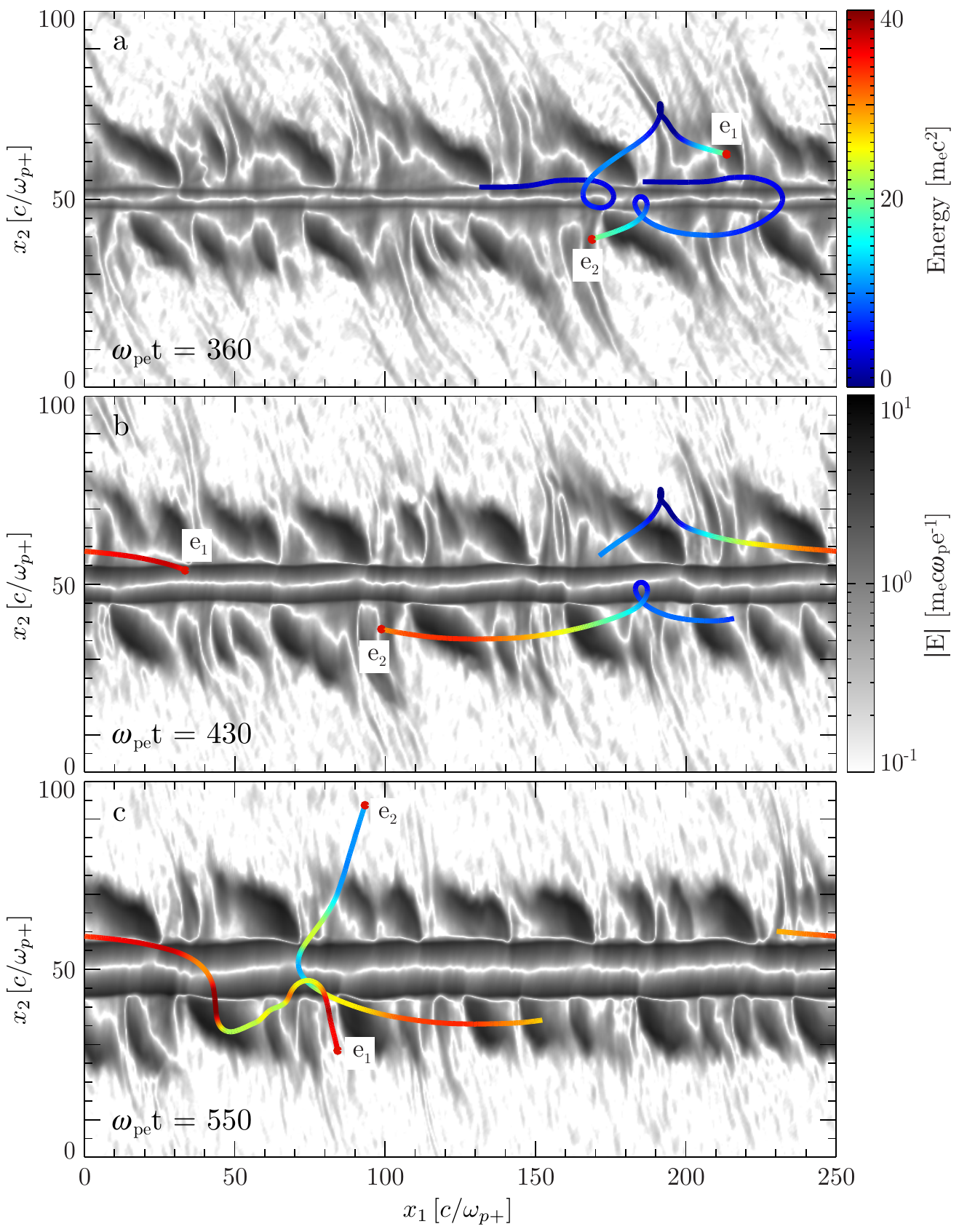}
\caption{Trajectories of two electrons during the non-linear phase of the instability. The varying color displayed along the trajectories stands for the energy of the electrons during the acceleration process. The black and white background represents the total electric field magnitude.}
	\label{fig:electraj}
\end{figure}

\begin{figure}[]
\centering
	\includegraphics[width=0.7\columnwidth]{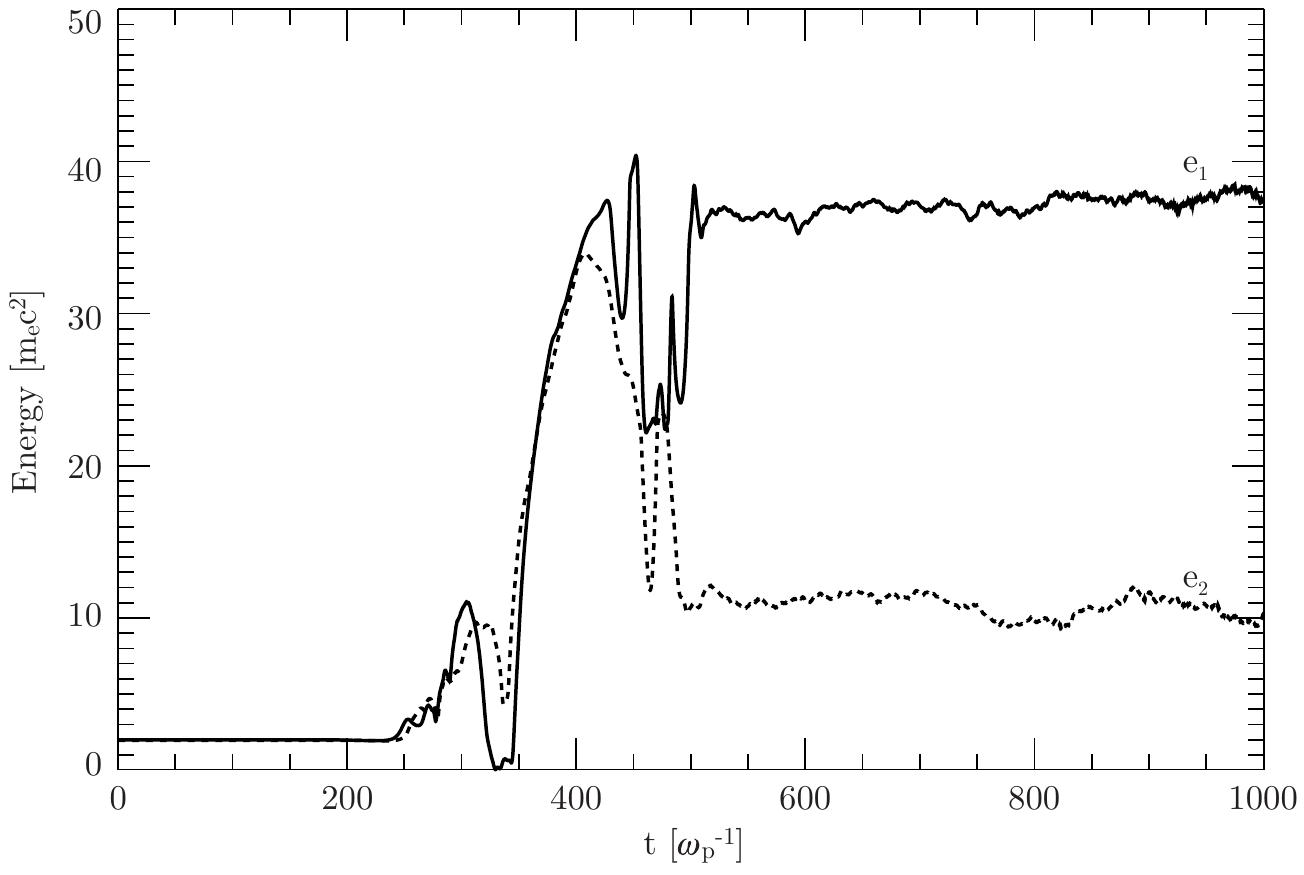}
	\caption{Time evolution of the energy of the same two electrons from Fig. \ref{fig:electraj}}
	\label{fig:enevstime}
\end{figure}

To gain deeper insight into the acceleration process of the most energetic electrons in the shear region, we followed various electron trajectories whose final energy lied in the second and third part of the distribution shown in Figure \ref{fig:gamdist}. Figure \ref{fig:electraj} shows segments (on every inset the electron is tracked during $\omega_pt\sim 100$) of two electron trajectories. The two electrons are initially in the vicinity of the shear, as shown in Fig. \ref{fig:electraj}-a. The black and white background represents the total electric field magnitude. The growing middle structure represents the dc part of the field whereas the modulated patches on both sides originate from the saturated unstable modes. One can clearly see that the most of the acceleration occurs when the electrons cross the electric field patches. The magnitude of the electric field patches is mainly due to the transverse component $E_{2}$, whereas the acceleration takes places in the $x_1$ direction. After being accelerated, the particle can cross the shear to be finally reaccelerated on the patches of the other side or definitively leave the shear region. The time evolution of the energy of the two tracked electrons is shown in Fig. \ref{fig:enevstime}. As explained previously, the energy of the electrons does no increase significantly until $\omega_pt\sim 300$ which corresponds to the saturation of the KHI. Soon thereafter both electrons experience a strong energy kick, $\Delta \gamma \sim 30$ at around $\omega_pt\sim 50$. At $\omega_pt\simeq 430$, the electrons have acquired their maximum energy that then remains constant, indicating they have left the field dominated shear region. The electron denoted by $e_1$ crosses the shear without really being affected by the transverse component of the electric field while the electron $e_2$ experiences a strong deceleration before eventually leaving the shear region. 

One can understand the acceleration mechanism if one sees the process as relativistic electrons surfing on the electric field patches, flowing with the plasma that, can be assumed constant in magnitude. First it is necessary to quantify the magnitude of the three electromagnetic components. Using Eq. (\ref{eq:curr_x}), Eq. (\ref{eq:maxwell1}) and Eq. (\ref{eq:maxwell4}), we obtain the following ratio in the limit $\gamma_0\gg 1$: $|E_{1}/E_{2}|\sim 1/\sqrt{2}\gamma_0$ and $|E_{2}/B_{3}|\sim 1$. If we assume that those ratio still hold during the non-linear phase, then the orbits of the electrons, approximatively streaming in the $x_1$ direction with the Lorentz factor of $\gamma_0$, are mainly governed by $E_{x2}$ and $B_{x3}$. The equations of motion read 
\begin{equation}
\frac{dp_{1}}{dt}=-\frac{e}{c}v_{2}B_{3},~\frac{dp_{2}}{dt}=-e\left(E_{2}-\frac{v_{1}}{c}B_{3}\right).
\end{equation}
In the case $E_2=B_3$, the solution for the trajectory of the electron is given by \cite{Landau}
\begin{eqnarray}
p_1 = -\frac{\alpha}{2c}+\frac{c^2p_2^2+\epsilon^2}{2\alpha c} \\
\frac{c^2}{3\alpha^2}p_2^3+\left(1+\frac{\epsilon^2}{\alpha^2}\right)p_2 = -2eE_2t,
\end{eqnarray}
with $\alpha = mc^2\gamma-cp_1$ and $\epsilon=mc^2$. One can clearly see that no matter what is the initial condition, for long enough time, $p_2/p_1 \sim 2\alpha /cp_2$ which means that the momentum increases most rapidly in the direction perpendicular to $E_2$ and $B_3$, i.e., $x_1$. This confirms the observations from the simulations where the acceleration is mostly directed towards the $x_1$ axis where the component of the electric field is transverse. The maximum energy gain an electron undergoes while encountering an electric field patch is given by
\begin{equation}
\Delta\mathcal{E} = emc^2\int_{patch} d\mathbf{x}\cdot\mathbf{E}\sim emc^2\int_{patch} dx_2{E}_{2}.
\end{equation}
We treat the electric field as constant and we assume that its magnitude is on the order of its saturation value, $E_\mathrm{sat}\sim \sqrt{\gamma_0}~m_e\omega_pc/e$. The maximum length of an electric field patch is typically $1/k_\mathrm{max}=\sqrt{8/3}\gamma_0^{3/2}c/\omega_p$. Since the patches are drifting with the initial flow, one must take into account the difference in speed between the electron being accelerated and the flow: $\Delta v = v_e - v_0$. Assuming $\gamma_e \gg \gamma_0 \gg 1$ ($\gamma_e$ is the Lorentz factor associated to the electron speed $v_e$), we obtain
\begin{equation}
\Delta\mathcal{E}_\mathrm{max} \sim E_\mathrm{sat}\frac{c}{k_\mathrm{max}\Delta v}\propto mc^2\gamma_0^4.
\end{equation}

For the parameters of our simulations, the cut-off of the spectra is typically on the order of $\gamma_0^4$ and in agreement with the energy spectra cut-off obtained in \cite{Liang213} .The energy kick experienced by the electrons around $\omega_p t=400$ in Fig.\ref{fig:enevstime} is about $\Delta\mathcal{E}\sim 40$ for both electrons, which is smaller than $\Delta\mathcal{E}_\mathrm{max}$ since they are encountering here smaller patches . As was previously illustrated with electrons $e_1$ and $e_2$, the energy kick does not determine the final energy of an electron before it leaves the shear region. In fact, depending on its transverse momentum, the electron can either experience another acceleration or a strong deceleration. This erratic feature allows to understand the wide range of energies among which the electrons are distributed in Fig. \ref{fig:gamdist}.

\newpage
\section{Conclusions}

Shear instabilities in plasmas are usually studied within the framework of magnetohydrodynamics where the plasma is considered as a magnetized fluid and where the typical time scale is governed by ion motion. We have shown in this work that electron scale physics leads to a variety of new effects when one considers an initially unmagnetized cold shearing collisionless plasma.

The collective electron dynamics can be described in first approximation by using a two-fluid model that allows to take into account electron inertial physics, not captured in MHD models.  In this fluid framework, we have presented the derivation of the equations for the linear development of the longitudinal KHI, assuming arbitrary velocity and density profiles. This framework was applied to a special case, where the velocity and density profiles were simple step-functions, allowing analytical solutions to the equations. The model provided new insights into the effect of density-contrasts between shearing flows, namely that  the development of the KHI is robust to density jumps, making it ubiquitous in astrophysical settings. We also observed that the unstable modes begin to drift when the density symmetry is broken. In large density-contrast regimes, the KHI dominates over other common astrophysical plasma instabilities such as the Weibel and Two-Stream instabilities. The case of a finite shear profile has also been investigated in detail. A smooth velocity profile (non step-like) induces a phase mixing of the eigenmodes of the system that results into a damping term. The combined effect of the instability due to the shear with the damping term  suggests that the maximum growth rate is a decreasing function of the shear gradient length. All of these results obtained in the limit of linearized fluid equations have been accurately verified by 2D PIC simulations. 

PIC simulation results have also demonstrated the emergence of a large-scale dc magnetic field after the onset of the electron-scale KHI . However this dc field is not captured by the two-fluid KHI theory nor MHD model. We have shown that the emergence of the dc magnetic field is intrinsically associated with electron-ion shear flows. The dc magnetic field is induced through the formation of dc current sheets driven by the expansion of electrons in the shear region due to either a thermal expansion or the development of the cold fluid electron-scale KHI perturbations. We have presented an analytical description of the formation of the dc field in agreement with 1D, 2D, and 3D PIC simulations. The dc magnetic-field saturation on the electron time scale is independent of the type of the expansion and persists beyond ion time scales.

Finally, we addressed the particle acceleration physics due to scattering in the self-generated fields of electron-scale instabilities triggered in unmagnetized shear flows. An understanding of how electrons are accelerated is essential if we are to fully interpret observations since it is presumably the radiation from energetic electrons that is most often observed from astrophysical sources. To address this issue, we have tracked the most energetic electrons in our simulations to identify the acceleration mechanism. It was found that the kinetic energy, initially stored in the drift, was mainly redistributed thermally. The bulk of the energy distribution energy of the electrons has a typical temperature comparable with the Lorentz factor of the flow. Nevertheless, the energy distribution also displays a non-thermal tail. The most energetic electrons, that make up the power-law tail of the distribution, are accelerated while surfing close to the speed of light on electric and magnetic field patches, self consistently developed by the electron-scale KHI, which are carried by the flow. This results in an efficient acceleration mechanism where electrons can reach energies on the order of $\gamma_0^4$, where $\gamma_0$ is the initial Lorentz factor of the flow.

\ack

E. P. Alves and T. Grismayer contributed equally to this work. This work was supported by the European Research Council (ERCÑ2010ÑAdG Grant 267841) and FCT (Portugal) grants SFRH/BD/75558/2010, SFRH/BPD/75462/2010, and PTDC/FIS/111720/2009. We acknowledge PRACE for providing access to resource SuperMUC based in Germany at the Leibniz research center. Simulations were performed at the IST cluster (Lisbon, Portugal) and SuperMUC (Germany).

\section*{Appendix: mathematical and numerical treatment of the effect of finite velocity shear gradient}
 \label{sec:appendix}

In this appendix we present the detailed mathematics and numerics underlying the study of the effect of a finite shear gradient on the development of the electron-scale KHI, i.e., non-step like transition regions between the two flows. The equation that is required to solve in order to calculate the effect of a finite shear gradient on the dispersion relation is Eq. (\ref{eq:eigenmodes}). For the sake of simplicity, we consider the case of a constant plasma density profile $(n_{+}=n_{-})$. For a uniform density profile, Eq. (\ref{eq:eigenmodes}) reads
\begin{equation}
		\label{eq:eigenmodes_smooth}
		\frac{\partial}{\partial x} \left [\epsilon \frac{\partial E_{y1}}{\partial x} \right ] - \kappa^2 E_{y1} = 0
	\end{equation}
where the functions $A$, $B$ and $C$ are now given by
	\begin{equation}
\cases{ 
			A=\epsilon=	\left(  \frac{1}{\gamma_0^2} \frac {\omega_p ^2}{\left( \omega -k v_0 \right) ^2} -1 \right)
			\\	
			B=0 \\	
			C=-\kappa^2=	-\epsilon k_{\perp}^2	
			}		
	\end{equation}
where $k_{\perp}^2=k^2+\omega_p ^2/c^2-\omega ^2/c^2$, $\omega=\omega_r+i\gamma$ and $v_0$ is a function of $x$. The boundary conditions are such that the field $E_{y1}$ should vanish at $|x|\rightarrow \infty$. It is useful to obtain an integral formulation of the latter equation in order to relate the conditions of instability of the system to the function $v_0$. To achieve this, let us first multiply Eq. (\ref{eq:eigenmodes_smooth}) by the conjugate of $E$ and integrate over $x$ to obtain
\begin{equation}
\label{int_formulation}
\int_{-\infty}^{\infty}dx\epsilon(x)\left(\left|\frac{\partial E}{\partial x}\right|^2+k_{\perp}^2|E|^2\right)=0
\end{equation}
that can also be written as 
\begin{eqnarray}
\label{real_int}
\int_{-\infty}^{\infty}dx\left(-1+\frac{\omega_p^2\left(-\gamma^2+(\omega_r-kv_0)^2\right)}{\left((\omega_r-kv_0)^2+\gamma^2\right)^2}\right)\left(\left|\frac{\partial E}{\partial x}\right|^2+k_{\perp}^2|E|^2\right)=0 \\
\label{imag_int}
\int_{-\infty}^{\infty}dx\frac{2i\omega_p^2\gamma(\omega_r-kv_0)}{\left((\omega_r-kv_0)^2+\gamma^2\right)^2}\left(\left|\frac{\partial E}{\partial x}\right|^2+k_{\perp}^2|E|^2\right)=0
\end{eqnarray}
by separating the real and the imaginary parts of Eq. (\ref{int_formulation}). If we look for an instability condition, one assumes that $\gamma>0$ and it is then clear that $(|\partial_x E|^2+k_{\perp}^2|E|^2) > 0$, which implies that in Eq.(\ref{real_int}) $-\gamma^2+(\omega_r-kv_0)^2$ should be positive somewhere in order for the integral to vanish. To satisfy the second condition, i.e, Eq.(\ref{imag_int}), we see that the function $\omega_r-kv_0$ cannot be strictly positive or negative. For instance, 
$\omega_r-kv_0$ can be an odd function. If $v_0$ is an odd function, then $\omega_r=0$. We will focus on the case of a symmetrical shear flow profile ($\omega_r=0$). Returning to the first condition, Eq.(\ref{real_int}), we see that $k^2v_0^2>\gamma^2$ somewhere and if $|v_0(x)|\leq V_0$ ($V_0$ the maximum absolute value of the flow velocity) and therefore the unstable modes should lie underneath the curves defined by $\gamma=\pm kV_0$. This is typically verified when one considers a tangential velocity shear, $v_0(x)=V_0\mathrm{sgn}(x)$, as in Section \ref{sec:disp_rel_analysis} where we obtained that $\Gamma\leq kV_0$.

If we return to Eq. (\ref{eq:eigenmodes_smooth}) and we expand the first term in Eq. (\ref{eq:eigenmodes_smooth}), the differential equation becomes 
\begin{equation}
\label{eq:eigenmodes_smooth2}
\frac{\partial^2 E_{y1}}{\partial x^2}+\frac{\partial \ln(\epsilon)}{\partial x}\frac{\partial E_{y1}}{\partial x}  - k_{\perp}^2 E_{y1} = 0.
\end{equation}

This differential equation can be encountered in various physical plasma scenarios. It has been extensively studied in the case electrostatic oscillation in a non uniform cold plasma by \cite{Barston64,Sedlacek71} and in the case of excitation of magnetic surface modes in MHD configurations by \cite{ChenHasegawa74,ZhuKivelson88}. In these previous works, the frequency spectrum and the eigenfunctions of the differential equation have been calculated and three main regimes were identified depending on the function $\epsilon$. When $\epsilon$ is an everywhere non-constant function, the spectrum is purely continuous and consists of those values of $\omega$ that satisfy the equation $\epsilon(x)=0$. In this case, the differential equation is always singular and the eigenfunctions can be constructed from well-known theorems on solutions of such equations near regular singularities. The well-behaved solution of the equation is obtained by an integral superposition over the whole spectrum of these eigenfunctions. On the other hand, as we already saw in Section \ref{sec:disp_rel_analysis}, a plasma with a discontinuous velocity profile and therefore a discontinuous dielectric constant $\epsilon$, exhibits different behavior. As a result of the analysis of \cite{Barston64}, the introduction of a jump discontinuity in $\epsilon(x)$ is necessary for the existence of well-behaved modes (so called surface modes) and the existence of a dispersion relation. The question that \cite{Sedlacek71} addresses is how to unite these two antagonistic pictures, which amounts to understand what happens to the surface wave when the velocity profile is smooth (or when the dielectric constant changes continuously from a region to another) and whether some collectives modes remain. 

If the velocity profile is odd and varies continuously from $-V_0$ to $V_0$, there is a local point in $x$ at which the shear flow satisfies the resonant condition $kv_0(x)=\omega_r=0$. At this point the spatial resonance in the smooth profile modifies the growth rate of the collective modes. One can interpret the modification of the growth rate by the introduction of damping and spatial dispersion to the originally surface wave of a discontinuous plasma. Physically, at the resonant point ,the eigenmodes face phase mixing which results in a wave damping. 

The theoretical framework to solve Eq. (\ref{eq:eigenmodes_smooth2}) was given by \cite{Sedlacek71}. Formally, one needs to construct the Green's function $G(x,x',\omega)$ of the differential equation where the mode solution of the electric field is given by
\begin{eqnarray}
\label{eq:formal_solution}
E_{y1}(x,\omega)&=&\int dx'G(x,x',\omega)E_{y1}^0(x',\omega) \\
E_{y1}(x,t)&=&\int_F d\omega'E_{y1}(x,\omega)e^{-i\omega t} 
\end{eqnarray}  
where $E_{y1}^0(x',\omega)$ denotes the initial perturbation and $F$ the integration path that requires time causality and that is deformed around the singularities. The time asymptotic solution of the field comes from the contribution due to the singularities of the Green's function. More precisely, the contributions near the isolated singularities correspond to the collectives modes while the contribution of the integration along the branch cuts leads to the continuous spectrum \cite{Sedlacek71}. The general theory to obtain the Green's function of the differential Eq. (\ref{eq:eigenmodes_smooth2}) relies on a theorem demonstrated by \cite{Friedman56}. The Green's function can be expressed in terms of two linearly independent solutions $\psi_1$ and $\psi_2$ of the homogeneous Eq. (\ref{eq:eigenmodes_smooth2}), with $\psi_1$ satisfying the boundary condition at $x=-\infty$ and $\psi_2$ at $x=+\infty$. The Green's function is written as 
\begin{equation}
G(x,x',\omega)=J^{-1}[\psi_1(x,\omega)\psi_2(x',\omega)H(x'-x)+\psi_2(x,\omega)\psi_1(x',\omega)H(x'-x)],
\end{equation}
where $H$ is the Heavyside function and $J$ the conjunct of the two solutions defined by
\begin{equation}
J(k,\omega)=\epsilon(x)\left[\frac{d\psi_2}{dx}\psi_1-\frac{d\psi_1}{dx}\psi_2\right],
\end{equation}
which is a function independent of the variable $x$ (see \cite{Friedman56} for the details). It is clear that the singularities of the Green's function $G$ are given by the values of $\omega$ for which $J=0$. To obtain the new dispersion relation, one needs to therefore obtain an analytical expression for $J$ and then find the zeros of this latter function. One problem that arises is that is usually unlikely that $\psi_1$ and $\psi_2$ can be expressed in terms of standard functions for a given dielectric constant $\epsilon(x)$, which poses difficulties in obtaining further analytical results. However, when the function $\epsilon(x)$ has a linear profile given by
\begin{equation}
\label{eq:eps_linear}
\epsilon(x) = 
\cases{ 
\epsilon_1=\left(  \frac{1}{\gamma_0^2} \frac {\omega_p ^2}{\left( \omega +k V_0 \right) ^2} -1 \right)   & $x < -L/2$\\ \\
\frac{\epsilon_2-\epsilon_1}{L}x+\frac{\epsilon_1+\epsilon_2}{2}  &$-L / 2< x < L / 2$ \\ \\
\epsilon_2=\left(  \frac{1}{\gamma_0^2} \frac {\omega_p ^2}{\left( \omega -k V_0 \right) ^2} -1 \right)  & $x > L/2$
}
\end{equation}
the eigenmode of this equation has been given already by \cite{Sedlacek71}. For $k_{\perp}L \ll 1$, where $k_{\perp}=k_{\perp}(L=0)$ as a first approximation, the dispersion relation is given approximately by
\begin{equation}
\frac{1}{\epsilon_1}+\frac{1}{\epsilon_2}-\frac{i\pi k_{\perp}L}{\epsilon_2-\epsilon_1}=0,
\end{equation}
which reduces to the prior dispersion relation Eq. (\ref{eq:disp_rel}) when $k_{\perp}L\rightarrow 0$. From this crude description of the evolution of the dielectric constant in the transition region, one gets that the maximum growth rate decreases linearly with the small parameter $k_{\perp}L$
\begin{equation}
\label{theoryGvsL}
\frac{\Gamma_\mathrm{max}}{\Gamma_\mathrm{max}^0}\simeq 1-\frac{\sqrt{3}}{8}\pi k_{\perp}L,
\end{equation}
where $\Gamma_\mathrm{max}^0/\omega_p = \sqrt{1/8}$ stands for the growth rate obtained for a discontinuous profile and where the second term in the r.h.s of Eq. (\ref{theoryGvsL}), $\sqrt{3}\pi k_{\perp}L/8$, can be interpreted as the damping factor arising from the finite shear gradient. The wavenumber corresponding to the maximum growth rate follows a similar trend by slightly decreasing when the parameter $k_{\perp}L$ increases. 

Although the basic physical effects of the smooth transition on the growth rate have been discussed above for small values of $k_{\perp}L$, one aims to calculate with greater accuracy the evolution of the maximum growth rate when the characteristic length of the velocity profile is varied. Since it is not possible to obtain analytical expression for $J$, a numerical solution of Eq. (\ref{eq:eigenmodes_smooth2}) is required. The underlying numerical algorithm can be outlined as follows. For a given wavenumber $k$, a guess value for $\omega$ is chosen. By following the evolution of $\omega$ as the characteristic shear gradient length is continuously varied, one can choose the analytical solution for the sharp shear case (equation (\ref{eq:gr_gruzinov})) as the initial guess value for $\omega$; the initial guess value for higher shear gradient length should then be based on the previously calculated value for a smaller shear gradient length. For a given velocity profile, we numerically solve the differential equation Eq. (\ref{eq:eigenmodes_smooth2}) where the dielectric constant is evaluated for $k$ and the guess value $\omega$. The numerical solution of the electric field is then injected into the integral equation Eq.(\ref{int_formulation}), where we look now for the value of $\omega$ such that the integral of Eq.(\ref{int_formulation}) vanishes. We then iterate the process until the value of $\omega$ converges.

\end{document}